\documentclass[preprint,pre,aps,onecolumn,superscriptaddress,floatfix]{revtex4-1}

\usepackage{graphicx}
\usepackage{amsmath}
\usepackage{amsfonts}
\usepackage{amssymb}
\usepackage{mathtools}
\usepackage{multirow}
\usepackage{bm}
\usepackage{pifont}
\usepackage{color}
\usepackage{float}
\usepackage{relsize}
\usepackage[update,prepend]{epstopdf}

\usepackage{color}

\newcommand{\la}{\left\langle}
\newcommand{\ra}{\right\rangle}

\def \figwidth {12cm}

\begin{document}
\title{Transport coefficients of self-propelled particles:\\Reverse perturbations and transverse current correlations}
\author{Arash Nikoubashman}
\affiliation{Institute of Physics, Johannes-Gutenberg-University Mainz, Staudingerweg 7, 55128 Mainz, Germany}
\author{Thomas Ihle}
\affiliation{Institute for Physics,
University of Greifswald, 17489 Greifswald, Germany}

\begin{abstract}
The reverse perturbation method [Phys. Rev. E {\bf 59}, 4894 (1999)] for shearing simple liquids and
measuring their viscosity is extended to the Vicsek-model (VM) of active particles [Phys. Rev. Lett.
{\bf 75}, 1226 (1995)] and its metric-free version. The sheared systems exhibit a phenomenon that
is similar to the skin effect of an alternating electric current: momentum that is fed into the
boundaries of a layer decays mostly exponentially towards the center of the layer. It is shown how
two transport coefficients, {\it i.e.} the shear viscosity $\nu$ and the momentum amplification
coefficient $\lambda$, can be obtained by fitting this decay with an analytical solution of the
hydrodynamic equations for the VM. The viscosity of the VM consists of two parts, a kinetic and a
collisional contribution. While analytical predictions already exist for the former, a novel 
expression for the collisional part is derived by an Enskog-like kinetic theory. To verify the
predictions for the transport coefficients, Green-Kubo relations were evaluated and transverse
current correlations were measured in independent simulations. Not too far to the transition to
collective motion, we find excellent agreement between the different measurements of the transport
coefficients. However, the measured values of $\nu$ and $1-\lambda$ are always slightly higher than
the mean-field predictions, even at large mean free paths and at state points quite far from the
threshold to collective motion, that is, far in the disordered phase. These findings seem to
indicate that the mean-field assumption of molecular chaos is much less reliable in systems with
velocity-alignment rules such as the VM, compared to models obeying detailed balance such as
Multi-Particle Collision Dynamics.
\end{abstract}

\pacs{87.10.-e,05.20.Dd,64.60.Cn,02.70.-c} 

\maketitle

{\small PACS numbers:87.10.-e, 05.20.Dd, 64.60.Cn, 02.70.Ns} 

\section{Introduction}
\label{sec:intro}
During the past two decades, there has been a large interest in active matter systems, such as bird
flocks \cite{animal_flocks}, swarming bacteria \cite{kearns_11, copeland_09}, active colloids
\cite{zoettl_14, ginot_15}, microtubule mixtures \cite{nedelec_02} and actin networks \cite{actin_net}
driven by molecular motors. These systems display interesting behaviors such as pattern formation,
collective motion and non-equilibrium phase transitions \cite{vicsek_12, marchetti_13}. Some of these
features already occur in one of the simplest models for active matter, the Vicsek-model (VM) of
self-propelled particles \cite{vicsek_95, czirok_97, nagy_07} and its variants \cite{peruani_08,
aldana_09, peng_09, barbaro_09, ginelli_10a, chou_12, mishra_12, romensky_14}. Because of the
simplicity of its interaction rules and the existence of a non-standard transition to a collective
state of polar order, the VM became an archetype of active matter.

Due to the many degrees of freedom, theoretical studies of active matter systems are often based on
coarse-grained macroscopic transport equations for the slow variables such as density or momentum. 
Originally, the general forms of these equations were postulated by symmetry and renormalization
group arguments, such as in the seminal Toner-Tu theory \cite{toner_95, toner_98, toner_12} for
polar active matter. However, this approach leaves the coefficients of the terms in the transport
equation largely undetermined. Furthermore, memory and other nonlocal terms are usually not
considered, although for particular models there is evidence on their relevance \cite{kuersten_17}.
These shortcomings motivated many researchers to derive macroscopic transport equations directly
from the microscopic interactions and to obtain explicit expressions for the occurring coefficients
\cite{bertin_06, baskaran_08a, baskaran_08b, bertin_09, ihle_11, peshkov_12, farrell_12, grossmann_13,
thueroff_13, hanke_13, ihle_14_a, peshkov_14_b, chepizhko_14}. Even though most of these
coarse-graining approaches are based on some type of mean-field assumption, they can still be rather
involved and, in addition, rely on further approximations such as time scale separation, the
thermodynamic limit or the irrelevance of higher order spatial gradients.

In principle, different kinetic theory approaches for the same microscopic model can lead to
different macroscopic expressions, see for example Ref.~\citenum{bertin_15}. Thus, the validity of
the derived transport coefficients and of the macroscopic description in general is often
questionable and has led to debates \cite{aldana_09, ihle_13,  peshkov_14_b_C, ihle_14_b, ihle_14_c}.
To the best of our knowledge, so far there has been no comprehensive work on the verification of
transport coefficients in polar active matter. In this article, we start filling this void, at least
for two transport coefficients of the standard and the metric-free (or topological) VM. In particular,
we perform extensive agent-based simulations of the VM, measure the kinematic viscosity, $\nu$, and
the momentum amplification coefficient, $\lambda$, by means of several complementary methods, and
compare them to predictions from kinetic theory. To test our numerical tools, we perform additional
measurements on a momentum-conserving, particle-based model for computational fluid dynamics, called
Multi-Particle Collision Dynamics (MPCD) fluid \cite{malevanets_99, malevanets_00, gompper_08,
kapral_08, howard_18, howard_19}.

First, we employ a non-equilibrium approach and measure the response to shear, generated through the
reverse perturbation (RP) method \cite{mueller_plathe_99}. When applying RP to the VM, one observes
a phenomenon that is similar to the skin effect of an alternating electric current: momentum that is
fed into the boundaries of a channel decays mostly exponentially towards the center of the channel.
We show how $\nu$ and $\lambda$, can be obtained by fitting this decay with an analytical solution
of the hydrodynamic equations for the VM. In analogy to the MPCD case, the viscosity of the VM
consists of two parts, the kinetic ($\nu_{\rm kin}$) and the collisional viscosity
($\nu_{\rm coll}$). The latter contribution was missing in previous theories of the VM, and we
derive here an analytical expression for $\nu_{\rm coll}$ by extending the Enskog-like kinetic
theory from Refs.~\citenum{ihle_11, ihle_16}.

Furthermore, we introduce and apply two other methods to evaluate $\nu$ and $\lambda$ -- the
transverse current correlation method (TC) \cite{hoheisel_88, palmer_94} and the Green-Kubo (GK)
method \cite{green_54, kubo_57, zwanzig_65, forster_75}, which, in contrast to RP, operate without
introducing velocity gradients. We discuss colored noise and estimate memory effects in the theory
for TC, which explains limitations of this method at very small mean free path. We find excellent
agreement between the measurements of the RP and the TC method. Reasonable quantitative agreement
between agent-based simulations (using the RP, GK and TC method) and predictions by kinetic theory
is observed. This supports previous concerns on the validity of the mean-field assumption of
molecular chaos in systems without detailed balance and underlines the need for a theory that
includes correlation effects.

The rest of the manuscript is organized as follows. In Sec.~\ref{sec:model} we present the standard
and metric-free version of the VM, and briefly discuss the RP method for applying shear flow. In
Sec.~\ref{sec:theory} we develop analytic expressions for the transport coefficients $\lambda$ and
$\nu$, and discuss the theory of the TC method. In Sec.~\ref{sec:results} we shows our numerical
results and compare them to the theoretical predictions. Finally, in Sec.~\ref{sec:sum} we provide
a final discussion of our results and briefly outline open questions. The Appendix contains
additional theoretical details related to the kinetic theory of the VM, as well as results for an
MPCD fluid under shear.

\section{Model and Methods}
\label{sec:model}

\subsection{The standard Vicsek model}
\label{sec:Vicsek}
The VM \cite{vicsek_95, czirok_97, nagy_07} consists of $N$ point particles at global number density
$\rho_0$, which move at constant speed $v_0$ in two dimensions. The positions and velocities of the
particles at time $t$ are given by ${\bf x}_i(t)$ and ${\bf v}_i(t)$, respectively. In the VM, the
particles are propagated {\it via} sequential streaming and collision steps with time step $\tau$.
(The term ``collision'' should not to be taken literally, but instead it just denotes any action that
changes the momentum of a particle.) During the streaming step, the particles move ballistically
\begin{equation}
	\label{STREAM}
	{\bf x}_i(t+\tau)={\bf x}_i(t)+\tau {\bf v}_i(t)\,.
\end{equation}
Because the speeds of the particles stay the same at all times, the velocities are parameterized by
the ``flying'' angles, $\theta_i$, {\it i.e.} ${\bf v}_i=v_0(\cos{\theta_i},\sin{\theta_i})$.

In the collision step, the directions $\theta_i$ are changed so that the particles align with their
neighbors within a fixed distance $R$ plus some external noise. In practice, a circle of radius $R$
is drawn around the focal particle $i$, and the average direction $\Phi_i$ of motion of the particles
within the circle is determined according to
\begin{equation}
	\label{COLLIS}
	\Phi_i={\rm Arg}\left( \sum_{\{j\}} {\rm e}^{i\theta_j} \right) 
\end{equation}
where the sum goes over all particles within the interaction range $R$ (including particle $i$).
Once all average directions $\Phi_i$ are known, the new directions follow as
\begin{equation}
	\label{VM_RULE}
	\theta_i(t+\tau)=\Phi_i+\xi_i ,
\end{equation}
where $\xi_i$ is the so-called angular noise. The random numbers $\xi_i$ are uniformly distributed 
in the interval $[-\eta/2,\eta/2]$, with noise strength $\eta$. The model uses parallel updating,
and in this paper we will also assume the so-called standard VM which uses a forward-updating rule.
Thus, the already updated positions ${\bf x}_i(t+\tau)$ are used for determining the average
directions $\Phi_i$ at time $t$.

Another relevant model parameter is the average particle number $M$ that can be found inside a
circle of radius $R$, {\it i.e.} $M=\rho_0 \pi R^2$ with the global number density $\rho_0$. The
dimensionless parameter $M$ measures the ratio of the interaction range $R$ to the average particle
distance $1/\sqrt{\rho_0}$. By increasing $M$ and/or decreasing the noise $\eta$, the VM can be
driven from a disordered phase to a phase of collective motion. The degree of alignment of the
particle velocities can be quantified through the polar order parameter
\begin{align}
	v_{\rm a} = \frac{1}{Nv_0}\left|\sum_{i=1}^N \mathbf{v}_i\right| .
	\label{eq:va}
\end{align}
Assuming a spatially homogeneous system, the threshold condition for this non-equilibrium phase
transition can be calculated within mean-field kinetic theory (see Appendix~\ref{sec:appendixA} for
details). For sufficiently small $M \ll 1$, the threshold noise $\eta_{\rm c}$ is predicted as
\begin{equation}
	\label{BVM_CRIT}
	\eta_{\rm c}=\sqrt{48 M\left({2\over \pi}-{1\over 2}\right)}\,.
\end{equation}
For parameters where the molecular chaos assumption is strongly violated, $\eta_{\rm c}$ can be much
lower than this theoretical prediction, sometimes by a factor between two and three. For more details
on the calculations and for a discussion of this transition, see Refs. \citenum{ihle_11, ihle_13,
solon_13, solon_14, ihle_16}.

\subsection{The Vicsek model with topological interactions}
\label{sec:topo}
Recent experiments by Ballerini {\it et al.} \cite{ballerini_08, cavagna_10} on flocks of starlings 
indicated that a Vicsek-like interaction rule with a fixed interaction range might not be appropriate
for animal flocks. Instead, a statistical analysis revealed that, on average, each bird interacts 
with a fixed number of neighbors, typically six to seven. This constitutes a topological or
metric-free interaction because the metric distance is not relevant; rather, it is a question of who
the closest neighbors are. Ballerini {\it et al.} argued further that, due to evolutionary pressure,
the main goal of interaction among individuals is to maintain cohesion. By comparing simulations
with the regular VM and a modified VM with metric-free interactions, they found that flocks, when
facing predators, kept cohesion much better in the metric-free model. These observations inspired
several other groups to study versions of the VM with topological interactions.

In this paper we will focus on a simple modification of the VM, which was suggested by one of us
\cite{chou_12}, because it allows an analytical description by a similar Enskog-like kinetic theory
as the one outlined in Appendix.~\ref{sec:appendixA}. In this model, the alignment rule of the
regular VM, given by Eqs. (\ref{COLLIS}) and (\ref{VM_RULE}), is slightly modified such that the
number of particles in every collision circle is kept constant and equal to $M$ at all times by
locally adjusting the interaction radius. Thus, only the $M-1$ closest neighbors together with
particle $i$ itself are included in the calculation of the average angle $\Phi_i$ of a particular
particle $i$. This procedure leads to large interaction ranges in areas with sparse populations,
whereas the interaction radius becomes small at locations with a high particle number density.
We introduce an effective interaction radius for this metric-free model, $R_{\rm eff}=\sqrt{M/(\pi \rho_0)}$,
which is always set to one by appropriately choosing the particle number density $\rho_0$.

In the regular VM, a larger local particle density leads to more robust alignment and stronger local
order. This behavior can be seen in the phase diagram of the VM, for example Fig.~1 in Ref. 
\citenum{ihle_11}. This coupling between density and order is the main reason behind the occurrence
of soliton-like density waves near the order/disorder threshold in the regular VM \cite{ihle_13}. In
the topological VM, however, density and order are decoupled because it is always the same number of
particles that participate in the alignment interaction. Therefore, the long-wave length instability
of the regular VM as well as the density waves are absent in the topological VM \cite{chou_12,
peshkov_12_b}.

\subsection{The reverse perturbation method}
\label{sec:reverse}
We performed non-equilibrium simulations to compute the shear viscosity from the simulations. These
approaches often provide significantly better signal to noise ratios compared to equilibrium methods,
such as the GK relation \cite{green_54, kubo_57, zwanzig_65, forster_75}. To generate shear flow in
our system, we employed the RP method \cite{mueller_plathe_99}, where the shear stress on the
system is imposed externally, by generating a momentum flux through a slab perpendicular to the flow
direction. This flux is achieved by swapping the particle velocities in the following way: first,
the periodic simulation box is subdivided into equally sized slabs with thickness $a$ along the
gradient direction of the flow ($y$). Then, particle $i$ in the $y=0$ slab with the largest positive
$e_x$ value and particle $j$ in the $y=L_y/2$ slab with the largest negative $e_x$ value are
identified, and their velocities are swapped. This swapping procedure artificially generates a
momentum flux, which gives rise to a physical flow.

If both particles have the same mass, as is the case in all our models, swapping conserves both the
linear momentum and the global kinetic energy. In our implementation, momentum swaps were applied to
the system with equal probability either before or after the collision step. Note, that when using
the RP method for the VM, it is crucial to also swap $e_y$ of the particle pair so that the particle
speed $v_0$ is conserved. We verified that this additional swapping does not introduce an unwanted
momentum flux in the $x$ direction.

\begin{figure}
	\includegraphics[width=7cm]{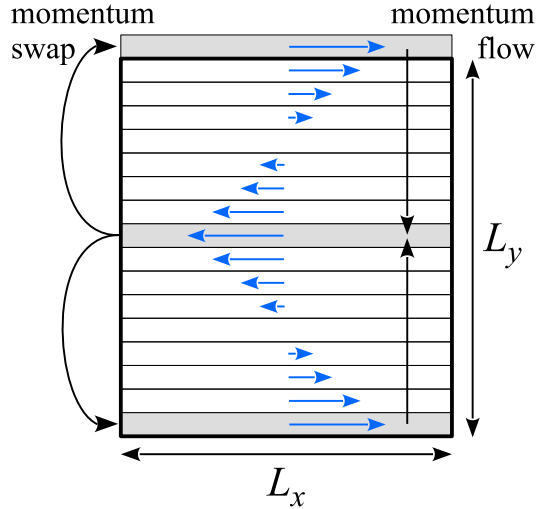}
	\caption{Schematic representation of the reverse perturbation method.}
	\label{fig:mpscheme}
\end{figure}

The imposed shear stress can be controlled by the amount of mome\textit{•}ntum swaps in one step and by the
time between swaps, $\Delta t$. For the chosen geometry of our two-dimensional systems, the average
shear stress can be computed as:
\begin{align}
	\langle \sigma \rangle = \frac{\langle \Delta p_x \rangle}{2\Delta t\, L_x} ,
\end{align}
where $\langle \Delta p_x \rangle$ is the $x$ component of the average total momentum exchanged
during one time step. Figure~\ref{fig:mpscheme} shows a schematic view of the shear procedure and
the emerging flow profile.

\section{Theory}
\label{sec:theory}
\subsection{The collisional viscosity $\nu_{\rm coll}$}
\label{sec:coll}
It has been shown by several groups \cite{tuzel_03, kikuchi_03, pooley_05, ihle_05, noguchi_08}, that
the kinematic shear viscosity of particle-based models, which consist of subsequent streaming and
collision steps, is a sum of two terms, namely the kinetic part, $\nu_{\rm kin}$, and the collisional
part, $\nu_{\rm coll}$,
\begin{equation}
	\label{SUM_VISC}
	\nu = \nu_{\rm kin}+\nu_{\rm coll}\,.
\end{equation}
Thus, it is plausible that such a decomposition is also valid for the VM. The kinetic part is due to
the momentum that is carried by a particle moving ballistically and can, for example, be calculated
by a Boltzmann-like kinetic equation. For the standard VM, this calculation has been done in Refs.
\citenum{ihle_11} and \citenum{ihle_16}, resulting in 
\begin{equation}
	\nu_{\rm kin}={v_0^2\tau\over 8}\,{1+p\over 1-p}\,.
	\label{KIN_VISC1}
\end{equation}
The auxiliary quantity $p$ involves an infinite sum,
\begin{equation}
	p = {4\over \eta} \sin{\eta} \sum_{n=1}^N {{\rm e}^{-M}\over n!}n^2 M^{n-1} K_{2c}^{11}(n),\,
	\label{P_DEF}
\end{equation}
where the coefficients $K_{2c}^{11}$ are given in Table I of Ref. \citenum{ihle_16}. Expression
(\ref{P_DEF}) can be evaluated approximately at small and large partner number $M$. For small 
normalized densities $M \ll 1$, a good approximation is
\begin{equation}	
	p \approx {\sin{\eta} \over \eta}\,{1+0.327M^2+0.072 M^3\over 1+M+M^2/2+M^3/6}\,.
	\label{P_SMALL_M}
\end{equation}
In the opposite limit, $M\gg 1$, one finds to leading order:
\begin{equation}
	p \approx {\sin{\eta}\over 2\eta}\,.
	\label{P_LARGE_M}
\end{equation}
In Ref.~\citenum{ihle_16} it was demonstrated that, like in regular fluids, the same expression for
$\nu_{\rm kin}$ can be obt`ined by evaluating a simple Green-Kubo relation by means of the molecular
chaos approximation.

The collisional contribution, $\nu_{\rm coll}$ in Eq.~(\ref{SUM_VISC}), stems from collisional
transfer of momentum across the finite interaction range $R$, and is therefore outside the scope of
Boltzmann-like equations. In contrast, an Enskog-like theory should be able to capture this
contribution (see Appendix~\ref{sec:appendixA}). However, in previous calculations \cite{ihle_11,
ihle_16}, a large mean-free path $\Lambda= v_0\tau \gg R$ was assumed, where $\nu_{\rm coll}$ becomes
negligible. In this manuscript, we show how to calculate $\nu_{\rm coll}$ for the standard VM within
mean-field kinetic theory. (Note, that Boltzmann approaches such as those of Refs. \citenum{bertin_06}
and \citenum{peshkov_14_a} are unable to obtain this important contribution to the viscosity.) The
details of this calculation have been moved to the Appendix~\ref{sec:appendixA} for conciseness, and
we summarize here only the final result for $\nu_{\rm coll}$
\begin{equation}
	\nu_{\rm coll}={R^2\over \tau} {\sin{(\eta/2)}\over 2 \eta}
	\sum_{n=1}^{\infty} {{\rm e}^{-M}\over (n-1)!} M^n\, K_C^1(n+1) .
	\label{NU_COLL_ALL}
\end{equation}
This term has been neglected in previous publications, and it is clear by dimensional analysis, that
the collisional part dominates the viscosity in the typical regime of the VM, such as originally
used by Vicsek {\it et al.} \cite{vicsek_95}, because $\Lambda \ll R$. This is because $\nu_{\rm kin}$
scales with time step $\tau$ and the effective temperature, $k_B T/m\sim v_0^2/2$, whereas
$\nu_{\rm coll}$ is proportional to $R^2/\tau$, thus $\nu_{\rm coll}/\nu_{\rm kin} \propto
(R/\Lambda)^2$.

\begin{figure}[htbp]
	\begin{center}
	\includegraphics[width=\figwidth]{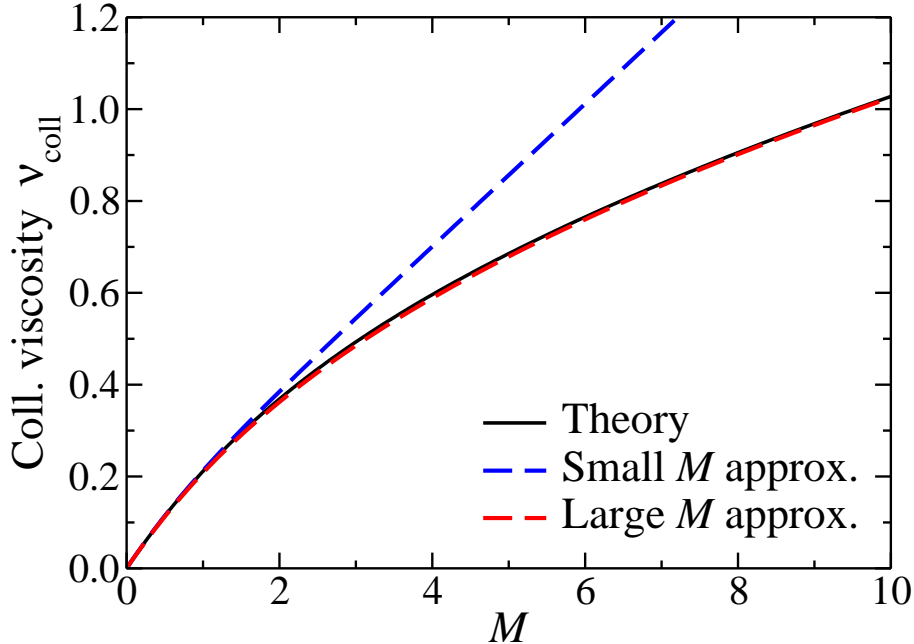}
	\caption{Collisional part of the kinematic shear viscosity, $\nu_{\rm coll}$, given by Eq.
	(\ref{NU_COLL_ALL}) (solid black line) {\it vs.} the normalized density, $M$. The blue dashed
	line shows the low density approximation, Eq. (\ref{NU_COLL_SMALL_M}), whereas the high density
	expression, Eq. (\ref{NU_COLL_LARGE_M}), is given by the dashed red line. The parameters are
	$\eta=3.2$, $\tau=0.2$, $R=1$, and $v_0=1$.}
	\label{compare_viscos_approx}
	\end{center}
\end{figure}

For $M \gg 1$, Eq. (\ref{NU_COLL_ALL}) can be approximated as
\begin{equation}
	\nu_{\rm coll} \approx {MR^2\over \tau}{\sin{(\eta/2)}\over 8 \eta}\sqrt{\pi\over M+2} ,
	\left[1+{3 M\over 8 (M+2)^2} \right]
	\label{NU_COLL_LARGE_M}
\end{equation}
by means of a saddle point expansion inside the infinite sum of Eq. (\ref{NU_COLL_ALL}). In the
opposite limit $M \ll 1$, we keep only the first terms in the sum and find 
\begin{equation}
	\nu_{\rm coll} \approx {MR^2\over \tau}{\sin{(\eta/2)}\over 2 \eta}\,\,
	\left[{1/\pi+0.2624 M+0.11245M^2+0.03347M^3\over 1+M+M^2/2+M^3/6}\right] .
	\label{NU_COLL_SMALL_M}
\end{equation}
Figure~\ref{compare_viscos_approx} shows the predicted collisional viscosity, Eq. (\ref{NU_COLL_ALL}),
as a  function of $M$ in comparison to the two approximations, Eqs. (\ref{NU_COLL_LARGE_M}) and
(\ref{NU_COLL_SMALL_M}). Interestingly, it turns out that the asymptotic expansion, 
Eq. (\ref{NU_COLL_LARGE_M}), is not only excellent for $M \ge 1$ but remains a very good approximation
for $M<1$ with an error of around one to two percent. In contrast, the approximative expression
Eq. (\ref{NU_COLL_SMALL_M}) which was obtained by truncating an infinite series becomes very
accurate at small $M$ but should not be used for $M>1$.

\begin{figure}[htbp]
	\begin{center}
	\includegraphics[width=\figwidth]{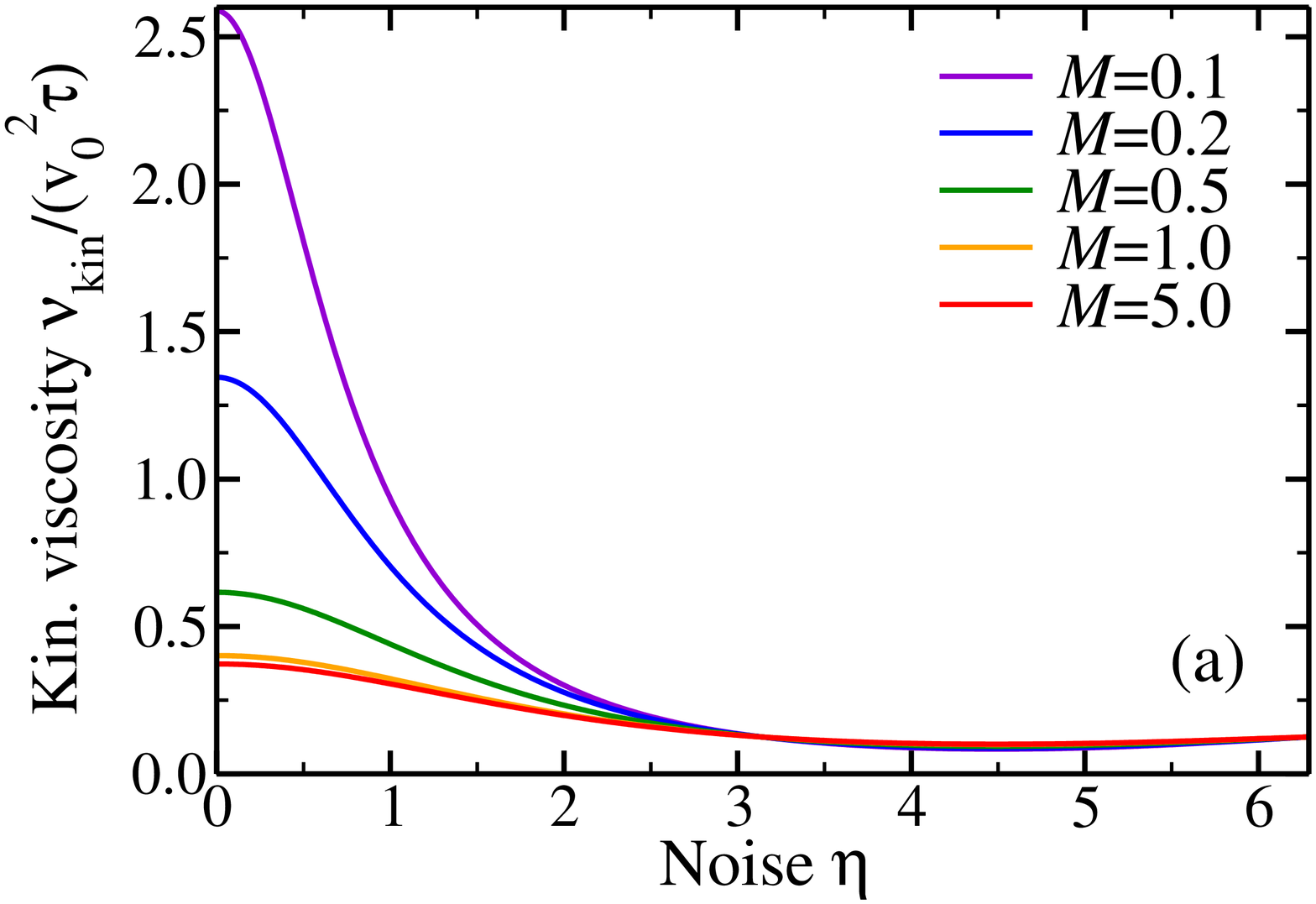}
	\quad
	\includegraphics[width=\figwidth]{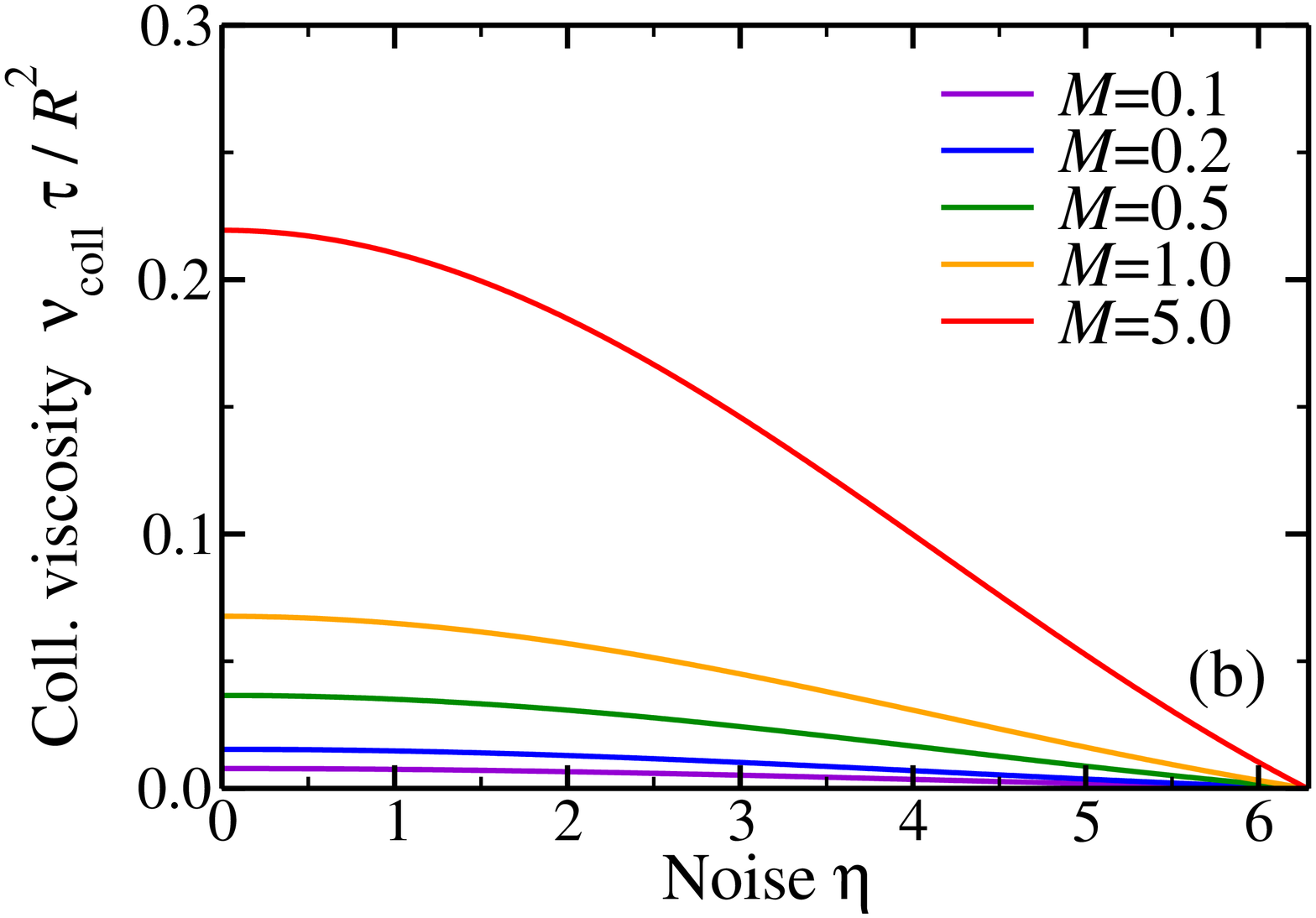}
	\caption{(a) Kinetic part of the kinematic shear viscosity, $\nu_{\rm kin}$, given by Eq.
	(\ref{KIN_VISC1}) and divided by $v_0^2 \tau$ {\it vs.} the noise, $\eta$, for various values of
	the dimensionless density, $M$. (b) Collisional part of the viscosity, $\nu_{\rm coll}$, from Eq.
	(\ref{NU_COLL_ALL}) and divided by $R^2/\tau$ {\it vs.} $\eta$.}
	\label{fig_kincoll_visc_vs_noise}
	\end{center}
\end{figure}

\begin{figure}[htbp]
	\begin{center}
	\includegraphics[width=\figwidth]{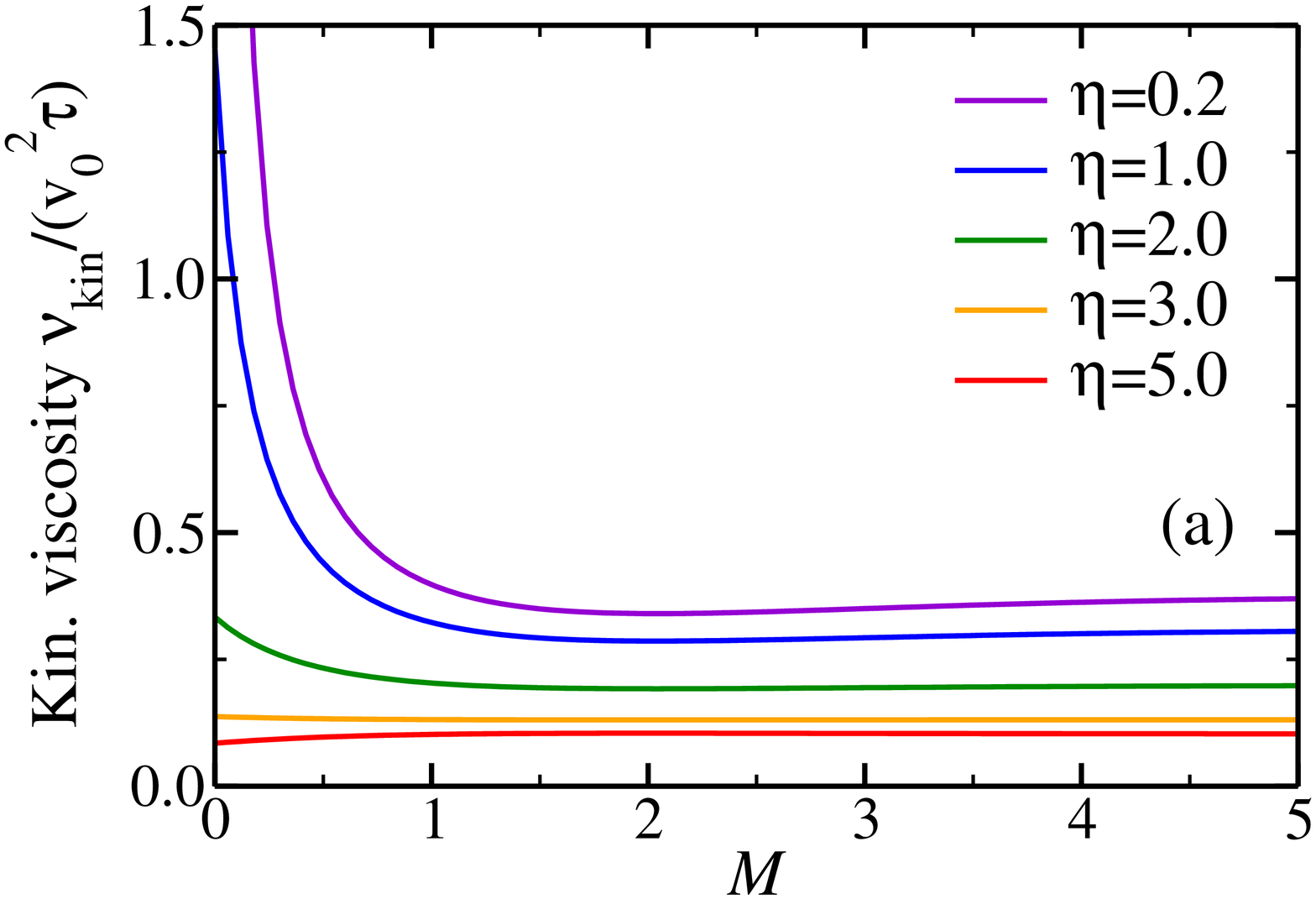}
	\quad
	\includegraphics[width=\figwidth]{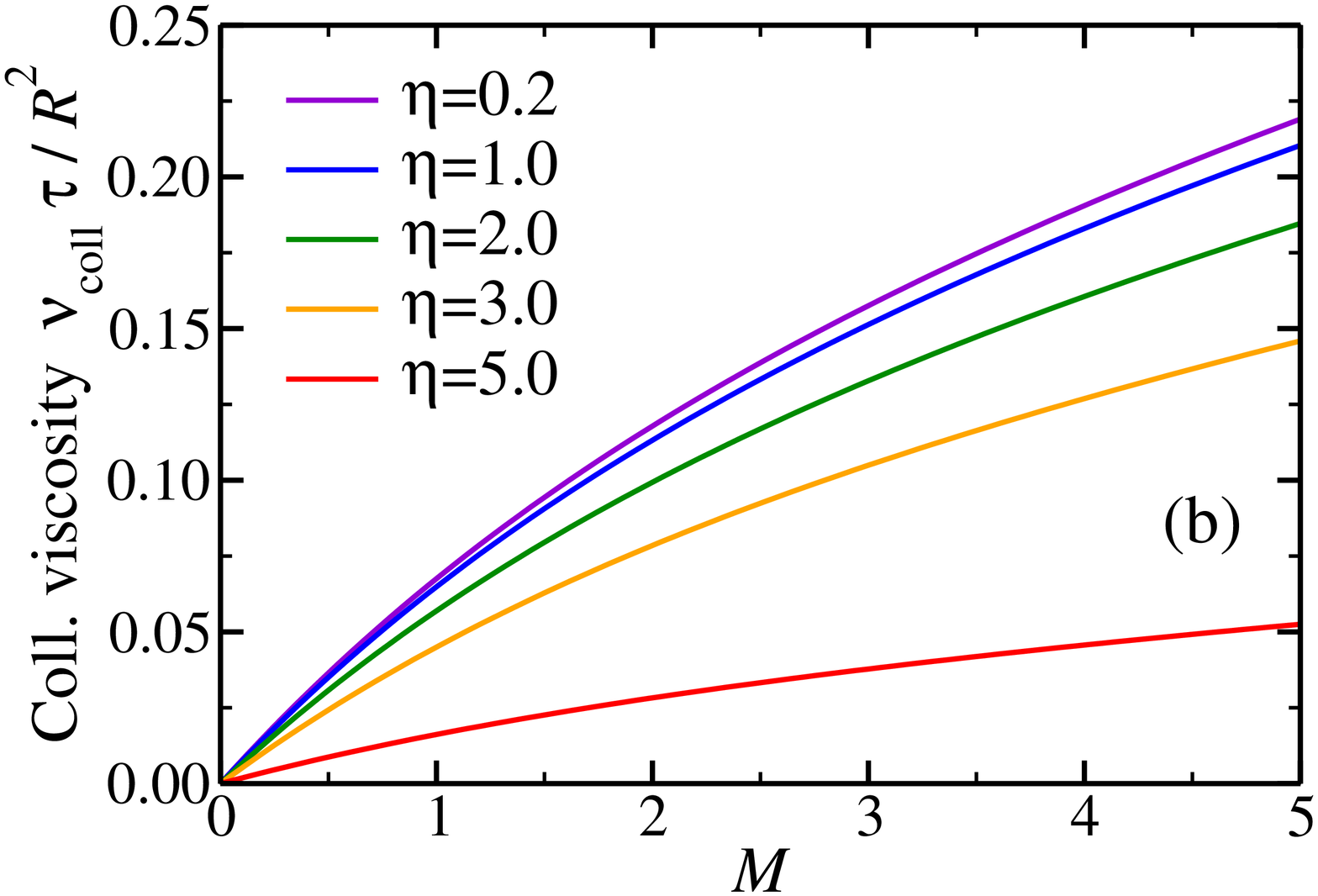}
	\caption{(a) Kinetic part of the kinematic shear viscosity, $\nu_{\rm kin}$, given by Eq.
	(\ref{KIN_VISC1}) and divided by $v_0^2 \tau$ {\it vs.} the dimensionless density, $M$, for
	various values of the noise, $\eta$. (b) Collisional part of the viscosity, $\nu_{\rm coll}$,
	from Eq. (\ref{NU_COLL_ALL}) and divided by $R^2/\tau$ {\it vs.} $M$.}
	\label{fig_kincoll_visc_vs_M}
	\end{center}
\end{figure}

\begin{figure}[htbp]
	\begin{center}
	\includegraphics[width=\figwidth]{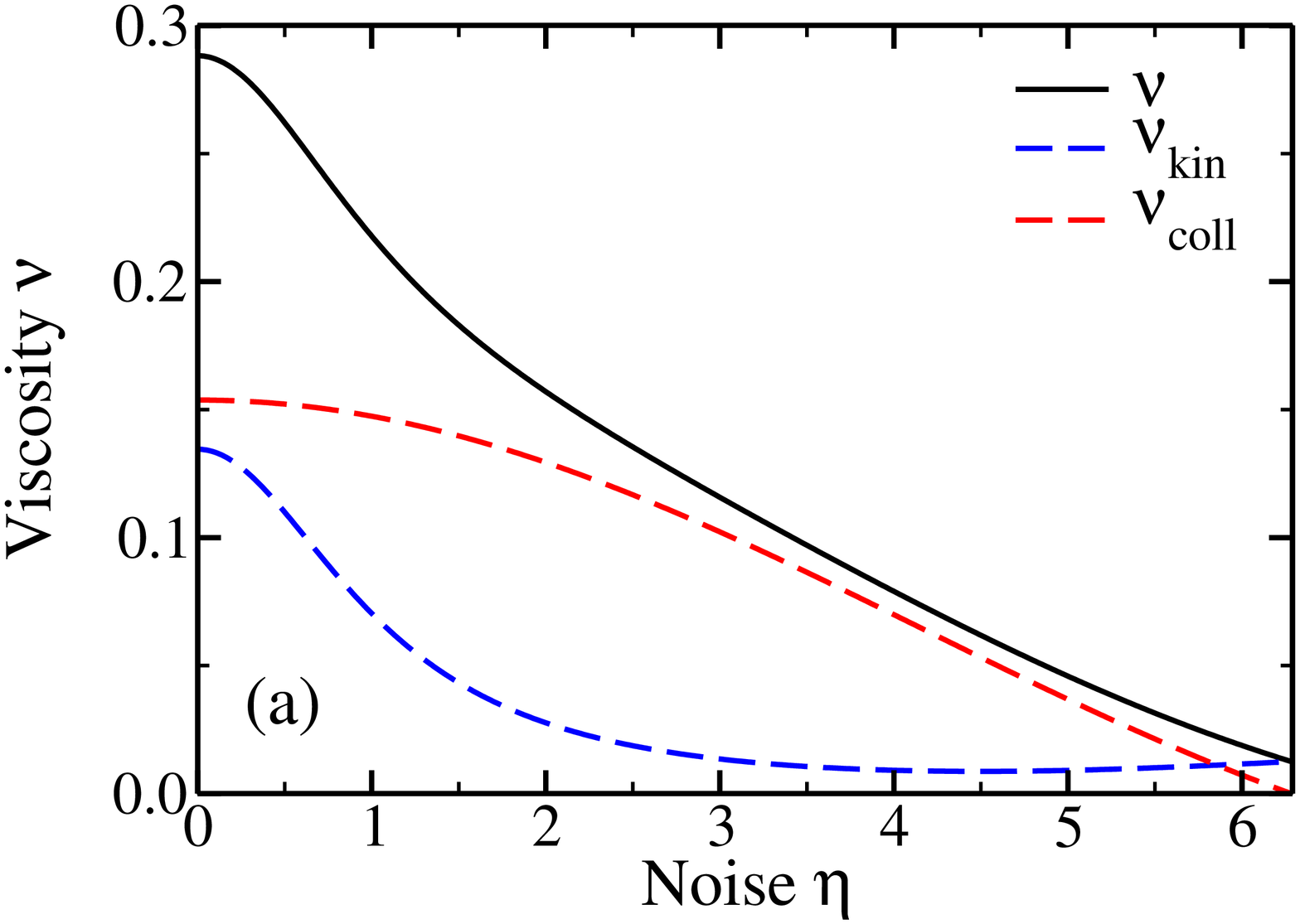}
	\quad
	\includegraphics[width=\figwidth]{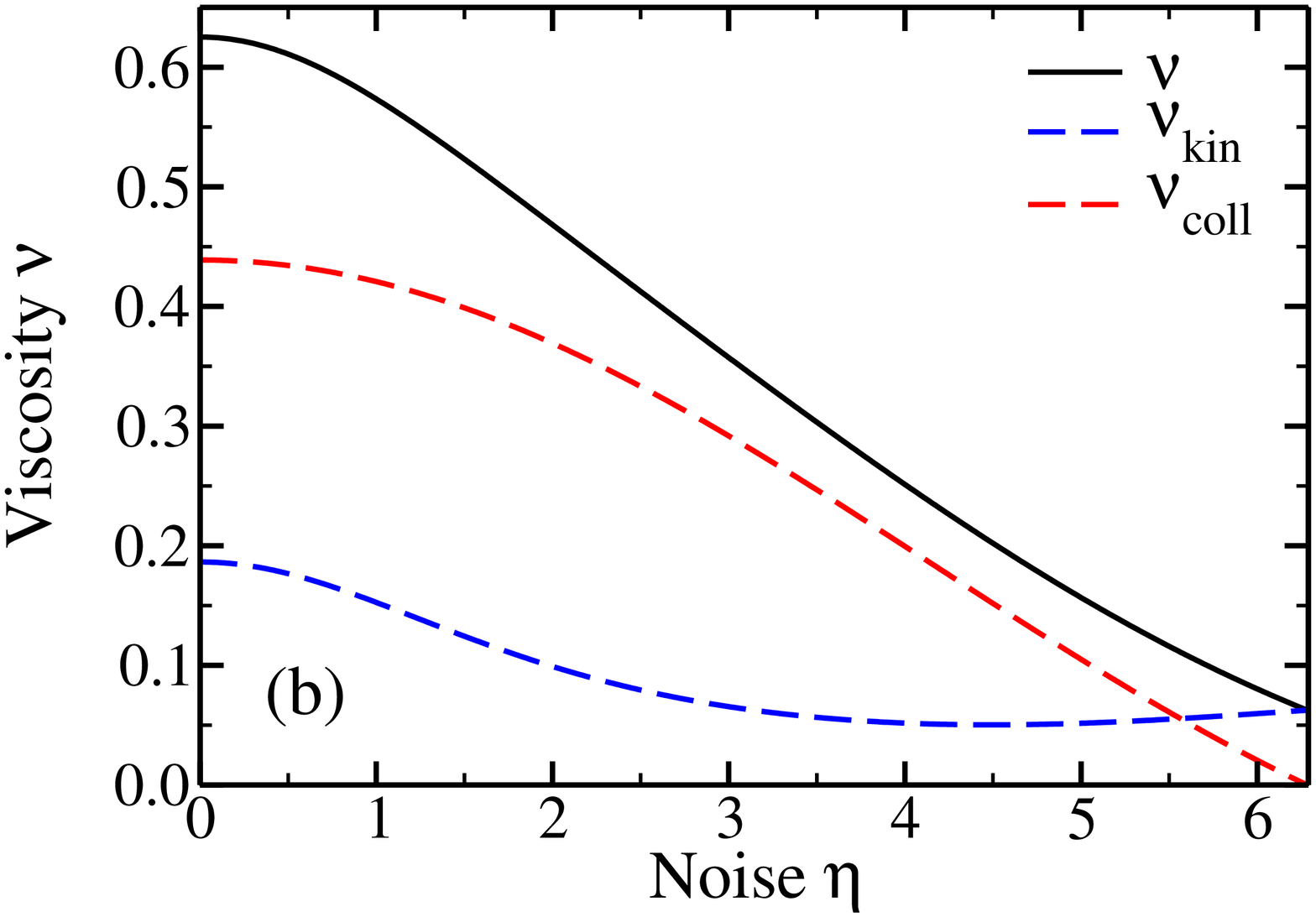}
	\caption{(a) Total shear viscosity, $\nu = \nu_{\rm kin} + \nu_{\rm coll}$ (solid line) given by
	Eqs. (\ref{KIN_VISC1}) and (\ref{NU_COLL_ALL}) {\it vs.} noise, $\eta$, for (a) $M=0.2$, $\tau=0.1$, 
	and (b) $M=5$, $\tau=0.5$. The dashed blue and red lines show the kinetic and collisional parts,
	$\nu_{\rm kin}$ and $\nu_{\rm coll}$, respectively, for comparison. Other model parameters are
	$R=1$ and $v_0=1$.}
	\label{fig_examples_visc_vs_noise}
	\end{center}
\end{figure}

Figures~\ref{fig_kincoll_visc_vs_noise} and \ref{fig_kincoll_visc_vs_M} show both (kinetic and
collisional) contributions to the viscosity as a function of noise, $\eta$, and normalized density,
$M$. The kinetic contribution is largest at both small $\eta$ and small $M$, whereas the
collisional contribution increases with $M$ and decreases with $\eta$.
Figure~\ref{fig_examples_visc_vs_noise} shows the total viscosity $\nu=\nu_{\rm kin}+\nu_{\rm coll}$
for two particular sets of parameters in comparison with $\nu_{\rm kin}$ and $\nu_{\rm coll}$.
Clearly, for these parameters, neglecting the collisional part leads to a large error.

Finally, we consider the system used in Vicsek's original paper, Ref.~\citenum{vicsek_95}. 
Translating the parameters from their Fig. 2(a) into our notation leads to $M=12.57$, $R=1$, and
$\Lambda=v_0 \tau=0.03$. Choosing $\eta=3.5$, which is slightly above $\eta_{\rm c}$, and applying
expressions (\ref{KIN_VISC1}) and (\ref{NU_COLL_ALL}), we predict $\nu_{\rm coll}=1.7$ and
$\nu_{\rm kin}=5.4\times 10^{-5}$. This finding confirms the expectation that the kinetic part 
of the viscosity is negligible here. Of course, these are predictions within the mean-field
approximation, which are not expected to be valid at this small ratio $\Lambda/R=0.03$. For improved
results, pre-collisional correlations as discussed in Ref. \citenum{chou_15} need to be taken into
account.
 
\subsection{Hydrodynamic theory for Vicsek-like models}
\label{sec:hydroTheory}
There is shared belief, that on a macroscopic level, polar active systems are described by a minimal
set of equations for mass and momentum density -- the well-known Toner-Tu equations \cite{toner_95,
toner_98, toner_12}. These equations were first postulated on the basis of symmetry and
renormalization group arguments. Within the mean-field assumption of molecular chaos, they have also
been derived from first principles, for a VM-like model with binary collisions by Bertin
{\it et al.} \cite{bertin_06, bertin_09}, and by Ihle \cite{ihle_11, ihle_14_a} for the
standard VM with discrete time evolution as considered here. In the latter approach, additional
nonlinear gradient terms which were not part of the original Toner-Tu theory, were found
\cite{ihle_16}.

We would like to apply the Toner-Tu theory to the shear setup given in Fig. \ref{fig:mpscheme}, in
order to extract the values of several transport coefficients from simulation data. Measurements are
taken in the stationary state, and thus, all time derivatives in the hydrodynamic equations are set
to zero. The stationary state is established by feeding a small amount of $x$-momentum into the top
layer of the channel, and by extracting $x$-momentum from the layer in the middle (see
Sec.~\ref{sec:reverse} for details). This procedure leads to a shear flow of small size at not too
low noise $\eta$. Therefore, we can neglect most nonlinear terms in the flow velocity. Furthermore,
there is no pressure gradient in the $x$-direction. Hence, we assume translational invariance for
that direction and neglect all spatial derivatives with respect to $x$. Because of the particular
way particle velocities are swapped, no net $y$-momentum is transferred between the feeding layers.
Analyzing the continuity equation, $\partial_t\rho+\partial_{\alpha} (\rho u_{\alpha})=0$, under
the previous assumptions shows that the transversal derivative, $\partial_y w_y$ of the $y$-component
of the momentum density, ${\bf w}=\rho{\bf u}=(w_x,w_y)$, should be zero. In shear flow of a regular
Newtonian fluid, the density is constant and the transversal velocity vanishes, $u_y=0$. This is
not quite the case for the active fluid considered here. Instead, due to a lack of Galilean invariance,
there are additional convective terms which prevent such a simple shear solution. Nevertheless,
agent-based simulations of the VM (see Sec.~\ref{sec:resultVM1}) showed that the density variations
across the channel are less than one percent, at least outside the parameter range where the
well-known density instability of the regular VM occurs \cite{bertin_06, ihle_11, ihle_13}.
Therefore, density gradients will be ignored in our theory.

Under these circumstances, the Toner-Tu equations for the components of the macroscopic velocity 
${\bf u}=(u_x,u_y)$ take a simplified form:
\begin{eqnarray}
\label{TONER1A}
\mu_1\, u_y\partial_y u_x+\mu_2\, u_x \partial_y u_y & \approx &(\nu \partial_y^2-\kappa -q u^2)u_x \\
\label{TONER1B}
(\mu_1+\mu_2+2\mu_3)\,u_y\partial_y u_x+2\mu_3\, u_x\partial_y u_x & \approx & 
(\nu \partial_y^2-\kappa -q u^2)u_y
 \,,\;\;{\rm with} \\
\nonumber
\kappa&\equiv&{1-\lambda\over\tau}\,.
\end{eqnarray}

The kinematic viscosity $\nu$, the coefficients of the convective terms, $\mu_1$, $\mu_2$, $\mu_3$,
and the strength of the cubic nonlinearity $q$ depend on the time step $\tau$, density $\rho$, noise
$\eta$ and the interaction radius $R$. The main difference to a regular fluid is the linear term in
${\bf u}$ which results from the violation of momentum conservation. The coefficient $\lambda$
describes, whether on average, momentum is lost or gained in a collision. The cubic term, $\propto
u^2{\bf u}$, becomes relevant below the threshold noise $\eta_{\rm c}$, where $\lambda>1$. The
threshold noise is defined by the condition $\lambda(\eta_{\rm c})=1$.

\subsubsection{Analysis for the disordered state, $\lambda<1$}
\label{sec:disord}
In the disordered state, $\eta > \eta_{\rm c}$, the momentum amplification factor $\lambda$ is
smaller than one, which means that, on average, momentum is lost in collisions. If momentum is
``fed'' into the boundary layer, it can only penetrate into the bulk of the channel within a certain
distance $l_S$ due to the interplay of momentum-diffusion and ``-evaporation''. Since this behavior
appears to be similar to the skin effect in electrodynamics, $l_S$ will be called skin depth. In
this scenario, we can neglect the cubic term in Eq. (\ref{TONER1A}). Since at $\eta > \eta_{\rm c}$
there is no spontaneous symmetry breaking, we also neglect the transversal component $u_y$. Both
assumptions have been justified numerically, and they allow us to obtain an analytical solution for
the velocity profile across the channel:
\begin{equation}
	u_x=d_0\,{\rm sinh}(d_1 y)\,,
	\label{V_PROFILE}
\end{equation}
where $u_x$ is the $x$-component of the macroscopic velocity. This profile is to be applied to the
upper (or lower) half of the channel with the $y$-coordinate set to zero in the middle of the
considered half-channel. The coefficient $d_1$ is given by
\begin{equation}
	d_1=\sqrt{1-\lambda\over \nu\,\tau}
	\label{A1EQ}
\end{equation}
As shown in Sec.~\ref{sec:resultVM1} of this paper, velocity profiles from agent-based simulations
which were averaged in time and over the length of the channel, show excellent agreement with this
${\rm sinh}$ profile. Fitting data to this profile enables the determination of the constants $d_0$
and $d_1$.

To recover both transport coefficients $\lambda$ and $\nu$, an additional quantity -- the momentum
flux -- is needed. The momentum flux $\sigma$ is determined by measuring the amount of momentum which
is fed into the top layer per time and length in the simulations. This flux is linked to the velocity
gradient by
\begin{equation}
	\sigma=\left.\nu\,\rho_m {\partial u_x\over \partial y}\right|_{y=L_y/4}
	\label{SIGMADEF}
\end{equation}

where $\rho_m$ is the mass density, and the gradient is to be evaluated at the top of the channel,
at $y=L_y/4$ ($y=0$ is defined in the middle of the upper half-channel, see Fig.~\ref{fig:mpscheme}).
Inserting the solution, Eq. (\ref{V_PROFILE}) into Eq. (\ref{SIGMADEF}) gives an equation for the
viscosity $\nu$:
\begin{equation}
	\nu={\sigma\over \rho_m d_0 d_1 {\rm cosh}(d_1 L_y/4)}
	\label{NU_FIND}
\end{equation}
Note, that using the coefficients $d_0$ and $d_1$ from a fit of the velocity profile, instead of
applying Eq. (\ref{SIGMADEF}) directly, circumvents the problem of numerically evaluating a velocity
gradient in the fluctuating top layer of the channel. Once $\nu$ has been determined, it can be
inserted in the relation for $d_1$, Eq.~(\ref{A1EQ}), yielding an expression for the coefficient
$1-\lambda$:
\begin{equation}
	1-\lambda=\tau d_1^2 \nu
	\label{LAM_FIND}
\end{equation}

It is possible to formally integrate Eq.~(\ref{TONER1A}) with $\mu_1=\mu_2=0$, but with the cubic
nonlinearity on the right hand side included. However, fitting this solution to numerically obtained
velocity profiles failed, in the sense, that it did not give reliable estimates for the coefficient
$q$. The reason is that the averaged velocities in our simulation data were too small for the
nonlinearity to be relevant. This was verified independently by using the mean-field prediction for
this coefficient from Refs. \cite{ihle_11, ihle_16}, evaluating the cubic term by hand and observing
that it is negligible compared to the linear terms.
 
\subsubsection{Analysis for the ordered state, $\lambda>1$}
\label{sec:order}
The situation in the ordered state is more complicated than the one at $\lambda<1$ because, (i) at
noise values slightly below the threshold noise, soliton-like density waves occur in the regular VM
\cite{chate_08, bertin_06, ihle_13}. That means, density gradients are large and derivatives with
respect to $x$ cannot be neglected. (ii) Spontaneous symmetry breaking occurs, leading to large 
macroscopic velocities that are not necessarily parallel to the walls of the channel. 

The former issue will be ignored, because there are models such as the metric-free VM
\cite{cavagna_10, ginelli_10a, chou_12} or the incompressible active liquid \cite{chen_15}, where
such density waves do not occur. Thus, for simplicity, in our analysis we still omit density
gradients. The latter issue means that nonlinear terms are relevant and that, depending on the
situation, the transversal component of the velocity $u_y$ could be larger than $u_x$. Here, as a
first step, we assume to be close to the threshold, $\lambda-1 \ll 1$. Furthermore, assuming small
velocity gradients and small momentum transfer rates, we still ignore the convective nonlinearities
but keep the cubic nonlinear term with coefficient $q$ to stabilize the solution. With these 
considerations, Eqs. (\ref{TONER1A}) and (\ref{TONER1B}) now become,
\begin{eqnarray}
	\label{TONER2a}
	\nu u_x^{''}-(\kappa +q u^2)u_x&=&0 \\
	\label{TONER2b}
	\nu u_y^{''}-(\kappa +q u^2)u_y&=&0\,,
\end{eqnarray}
where the definition $u_{\alpha}'\equiv {\partial u_{\alpha}\over \partial y}$ was used. Multiplying
the first equation (\ref{TONER2a}) by $u_x'$, the second one by $u_y'$, and adding both equations
yields:
\begin{equation}
	\label{CONSERVATION1}
	0={\partial \over \partial y}\left[-{\kappa\over 2}u^2-{q\over 4} u^4+{\nu\over 2}
	\left({\partial {\bf u}\over \partial y}\right)^2\right] .
\end{equation}

We define the flow velocity of a homogeneous ordered state, $u_0=\sqrt{|\kappa|/q}$, and the normal
vector $\mathbf{\hat{n}}$ for its flow direction. Near the threshold to collective motion,
$\lambda-1\ll 1$, and for a particular constant direction $\mathbf{\hat{n}}$, using
Eq. (\ref{CONSERVATION1}), we expand the solution around the homogeneous ordered state, and obtain
the approximate result
\begin{eqnarray}
	\label{SOL_FLUCT}
	\mathbf{u} & \approx & u_0\mathbf{\hat{n}}+A\,\mathbf{\hat{t}}\,  {\rm sinh}(d_1 y)\,\;\;{\rm with} \\
	\label{A1_ORDER}
	d_1&=&\sqrt{2(\lambda-1)\over \tau \nu}\,,
\end{eqnarray}
where the unit vector $\mathbf{\hat{t}}$ and the constant $A$ are arbitrary. In finite, not too
large systems, both directions $\mathbf{\hat{n}}$ and $\mathbf{\hat{t}}$ fluctuate over time. Because
our simulation data are time-averaged, such an average is also performed over Eq. (\ref{SOL_FLUCT}).
For the $x$-component of the flow velocity, one obtains
\begin{equation}
	\la u_x \ra = d_2+d_0\,{\rm sinh}(d_1 y) .
	\label{PROFILE_ORDER}
\end{equation}
Apart from the constant $d_2$, the solution has the same form as the one in the disordered state.
Note, however, that the coefficients $d_0$ and $d_2$ originate from the time-average of the
fluctuating unit vectors, $d_0 \equiv \la A\, \hat{t}_x \ra$, $d_2\equiv \la u_0\, \hat{n}_x \ra$, and 
therefore strongly depend on the system parameters and the details of the time-average. Thus, for
example, it is possible to observe an averaged flow profile with $d_2 \approx 0$ which deceivingly
looks like the one found in the disordered phase, even though particles have strong orientational
order at any given time. The procedures and formulas to obtain the viscosity, Eq. (\ref{NU_FIND}),
for both the ordered and the disordered phase are identical. However, there is a difference between
Eqs. (\ref{A1EQ}) and (\ref{A1_ORDER}) for the fit parameter $d_1$. Because of that, for
$\eta < \eta_{\rm c}$ one finds,
\begin{equation}
\label{LAM_FIND_ORDER}
	\lambda-1={\tau d_1^2 \nu\over 2}\,.
\end{equation}

\subsection{Transverse current fluctuations}
\label{sec:transverse}
The most common methods of calculating the shear viscosity from simulations are the GK approach and
nonequilibrium Molecular Dynamics. A third, less popular, approach is the use of transverse-current
auto-correlation functions. This method relies on the fact that in molecular liquids in thermal
equilibrium, long-wavelength fluctuations in transverse momentum fields decay exponentially with a
decay constant $\nu k^2$, where $\mathbf{k}$ is the wave vector of the fluctuation, and $\nu$ is
the kinematic shear viscosity. This approach was used, for example, to calculate the shear viscosity
for mono-atomic liquids in Ref.~\citenum{hoheisel_88}, and for liquid carbon dioxide, a molecular
fluid, in Ref.~\citenum{palmer_94}. In an ``artificial'' fluid without momentum conservation such as
the VM, the decay constant should contain an additional term which does not depend on $k$, because
even at zero wave number, momentum fluctuations still decay. This additional term should contain
information about the momentum amplification factor $\lambda$.

To describe small fluctuations in a stationary state, we start with the linearized Toner-Tu
equations for the momentum density ${\bf w}=(w_x,w_y)$ which, in the spirit of Landau-Lifschitz's
theory on ``fluctuating hydrodynamics'' \cite{landau_59}, are augmented with a noise source
${\bf \Psi}=(\Psi_x,\Psi_y)$,
\begin{equation}
	\label{START_TRANS1}
	\partial_t w_{\alpha}=-\partial_{\alpha} P-\kappa\, w_{\alpha} +
	\partial_{\beta}\sigma_{\alpha\beta}+\Psi_{\alpha} .
\end{equation}
Here, $P$ is the pressure, $\kappa\equiv(1-\lambda)/\tau$, and $\sigma_{\alpha\beta}$ is the viscous
stress tensor,
\begin{equation}
	\label{SIGMA_DEF}
	\sigma_{\alpha\beta}=\nu
	\left(\partial_{\alpha} w_{\beta}+\partial_{\beta} w_{\alpha}-{2\over d}
	\delta_{\alpha\beta}\partial_{\gamma} w_{\gamma}\right) ,
\end{equation}
where $d=2$ is the spatial dimension. The bulk viscosity is not included in Eq.~(\ref{START_TRANS1})
because it is irrelevant for the TC fluctuations. Higher order gradient terms and nonlinear terms
were neglected in this equation. We also assume that we are in the disordered phase, that is
$\kappa \geq 0$. As usual, the average of the noise can be chosen to vanish $\la {\bf \Psi} \ra=0$.
However, not much is known about its correlations; in general, 
we can neither assume that the noise is white nor
that its components are uncorrelated.

Defining the Fourier transform of the momentum density,
\begin{equation}
	\label{FOURIER_DEF_W}
	\hat{w}_{\alpha}({\bf k},t)=\frac{1}{V}\int w_{\alpha}({\bf x},t)\,{\rm e}^{i{\bf k}\cdot {\bf x}}\,{\rm d}{\bf x} ,
\end{equation}
Eq. (\ref{START_TRANS1}) reads in Fourier space as
\begin{equation}
	\partial_t\hat{w}_{\alpha}=ik_{\alpha} \hat{P} - \kappa\, \hat{w}_{\alpha} - 
	\nu k_{\beta}^2\, \hat{w}_{\alpha} + \hat{\Psi}_{\alpha} ,
	\label{TRANS_EQ_FOURIER}
\end{equation}
where $\hat{P}$ and $\hat{\Psi}_{\alpha}$ are the Fourier transforms of pressure and noise,
respectively. The simplest way to model a colored noise is to assume an  exponential
form
\begin{equation}
	\label{NOISE_CORR}
	\la \hat{\Psi}_{\alpha}({\bf k}, t) \hat{\Psi}^*_{\beta}({\bf k}, \tilde{t}) \ra =
	\frac{\gamma}{2} g_{\alpha\beta}({\bf k})\, C({\bf k})\,{\rm exp}(-\gamma|t-\tilde{t}|) ,
\end{equation}
where $C$ models the unknown (but irrelevant) strength of the noise and $\gamma$ is the decay rate
of the noise correlations. This leads to the definition of the memory time of the noise,
$\tau_{\rm N} \equiv 1/\gamma$. The tensor $g$, in particular its off-diagonal elements, describe
possible correlations between the different spatial components of the noise. For simplicity and for
symmetry reasons, one has $g_{xx}= g_{yy}=1$, and $g_{xy}=g^*_{yx}$. Without those correlations,
{\it i.e.} with $g_{xy}=0$, and in the limit $\gamma \to \infty$, the white noise behavior of a
regular fluid is recovered, where the correlations become equal to $\delta_{\alpha\beta} C
\delta(t-\tilde{t})$. All three quantities $\gamma$, $C$, and $g$ are likely to depend on the wave
vector $\mathbf{k}$. 

A general result of the Mori-Zwanzig projector operator formalism \cite{mori-zwanzig} is, that the
correlations of the (internal) noise are equal to the memory kernel in the corresponding generalized 
Langevin-Equation. However, our Langevin-equation, Eq. (\ref{TRANS_EQ_FOURIER}), contains only local
terms.  Therefore, even though we did not apply this formalism explicitly, for consistency
\cite{FOOT_MORI} we assume a white noise, {\it i.e.} $\gamma \rightarrow \infty$ in the following
calculations followed by an estimate of when this assumption is likely to fail. 

One way to extract the TC fluctuations is to focus on the vorticity $\bm{\omega} \equiv \nabla
\times {\bf w}$ of the flow, whose $z$-component is given by
\begin{equation}
	\label{VORTIC2}
	\omega_z=\partial_x w_y-\partial_y w_x .
\end{equation}
In Fourier space, Eq.~(\ref{VORTIC2}) becomes
\begin{equation}
	\label{OMEGA_DEF}
	\Omega\equiv \hat{\omega}_z=-ik_x \hat{w}_y+ i k_y \hat{w}_x\,.
\end{equation}
Multiplying the $x$-component of Eq. (\ref{START_TRANS1}) by $k_y$ as well as multiplying the
$y$-component by $k_x$, and subtracting both equation from each other leads to a closed equation for
$\Omega$,
\begin{equation}
	\label{STOCH_DGL}
	\partial_t \Omega=-\mu\, \Omega +\hat{\phi} ,
\end{equation}
where
\begin{eqnarray}
	\nonumber
	\mu & \equiv & \kappa+\nu k^2 \\
	\label{MU_DEF}
	\hat{\phi} & \equiv & ik_y \hat{\Psi}_x-ik_x \hat{\Psi}_y\,.
\end{eqnarray}
Thus, by using the vorticity, one has managed to effectively ``project out'' the longitudinal modes
which contain pressure and bulk viscosity. The correlations of the noise $\hat{\phi}$ follow from
Eq.~(\ref{NOISE_CORR}) in the limit $\gamma\rightarrow \infty$ as
\begin{equation}
	\label{VORT_NOISE_CORR}
	\la \hat{\phi}({\bf k}, t)\hat{\phi}^*({\bf k}, \tilde{t})\ra =
	G({\bf k})\,\delta(t-\tilde{t}) ,
\end{equation}
where
\begin{equation}
	G({\bf k})=C({\bf k})\,(k^2-k_x k_y[g_{xy}+g_{xy}^*]) .
\end{equation}
The stochastic differential equation, Eq. (\ref{STOCH_DGL}), is solved, and in the stationary limit,
$t \to \infty$ and $\tilde{t} \to \infty$, we obtain the vorticity correlations
\begin{equation}
\label{OMEGA_CORR1_W}
	\la	\Omega({\bf k},t)\Omega^*({\bf k},\tilde{t})\ra = 
\frac{G}{2\mu}{\rm exp}(-\mu\,|t-\tilde{t}|)\,.
\end{equation}

Let us define the viscosity-related decay time $\tau_{\mu}\equiv 1/\mu $ and the ratio of the two
characteristic times scales,
\begin{equation}
	\label{DEF_RATIO}
	\delta \equiv \frac{\tau_{\rm N}}{\tau_{\mu}} = \frac{\mu}{\gamma}\,.
\end{equation}
Once the decay time $\tau_{\mu}$ and the memory time of the noise $\tau_{\rm N}$ are approximately
equal, memory effects should matter for the decay of the vorticity correlations, and we expect
deviations from the white noise prediction, Eq. (\ref{OMEGA_CORR1_W}). 

In the collision-dominated regime of the VM, where $v_0\tau \ll R$, $\nu_{coll}/\nu_{kin} \gg 1$ and
$\nu \propto 1/\tau$, the total viscosity, $\nu$, can become very large at small time steps $\tau$.
Thus, $\mu$ would also become large. This trend to small $\tau_{\mu}=1/\mu$ is intensified if one is
deep in the disordered phase, where $\kappa$ is large too, and also in systems with a small linear
dimension $L$ since the smallest useful wavenumber is equal to $2\pi/L$ and thus can be rather large. 
In the same limit of small time steps, a particle needs more iterations to move out of the collision
circle of current collision partners. Hypothesizing, that, at least far in the disordered phase, the
memory time of the noise $\tau_{\rm N}$ is approximately given by the time two particles diffuse away
from each other by a distance $R$, we obtain a rough estimate for $\tau_{\rm N}$,
\begin{equation}
	\tau_{\rm N} \approx \frac{R^2}{2 v_0^2 \tau}
\end{equation}
For small time steps, we also have $\mu \propto 1/\tau$, and therefore 
\begin{equation}
	\delta = \mu \tau_{\rm N} \sim \left({R\over v_0 \tau}\right)^2 .
\end{equation}
Hence, we predict that at sufficiently small mean free path $v_0\tau$ (compared to the radius of the 
interaction circle), the ratio of the time scales, $\delta$, will become larger than 1, meaning that
the memory time of the noise cannot be neglected. Therefore, we apply the TC method only at
sufficiently large mean free paths where we can assume decent accuracy of Eq. (\ref{OMEGA_CORR1_W}). 
The corresponding numerical results are presented in Sec.~\ref{sec:numtrans}

\section{Numerical results}
\label{sec:results}
\subsection{Velocity profiles, polar order, and transverse current fluctuations}
\label{sec:resultVM1}
We performed two-dimensional agent-based simulations of the regular and the metric-free (topological)
VM with system sizes ranging from $16 \times 16$ to $128 \times 32$, and followed the RP protocol
with swap-times between $\Delta t=1$ an $\Delta t=4$ (see Sec.~\ref{sec:reverse}). The simulations
usually ran for $5 \times 10^6$ to $3 \times 10^7$ iterations after the stationary state has been
reached in order to ensure sufficient accuracy in the time averages. Since momentum is not conserved
in the VM, the stationary velocity profile across each of the half channels (see Fig.~\ref{fig:mpscheme})
is usually not linear. Hence, unlike as for momentum-conserving fluids such as MPCD (see Appendix
\ref{sec:appendixB}), the shear viscosity cannot simply be obtained as
the proportionality factor between the measured velocity gradient and the applied shear stress.
Instead, the theory outlined in Sec.~\ref{sec:hydroTheory} is used to evaluate the simulation data.
In particular, the velocity profile was fitted with a ${\rm sinh}$ profile, according to Eq.
(\ref{PROFILE_ORDER}). The extracted fitting coefficients $d_0$ and $d_1$ were inserted in Eq.
(\ref{NU_FIND}) to obtain the viscosity $\nu$. If the polar order parameter was above about $0.15$
and the coefficient $d_2$ significantly deviated from zero, expression (\ref{LAM_FIND_ORDER}) for
the ordered state was used to obtain the momentum gain coefficient $\lambda$. Otherwise, Eq.
(\ref{LAM_FIND}) for the disordered state was applied to extract $\lambda$.

\begin{figure}[htb]
	\includegraphics[width=\figwidth]{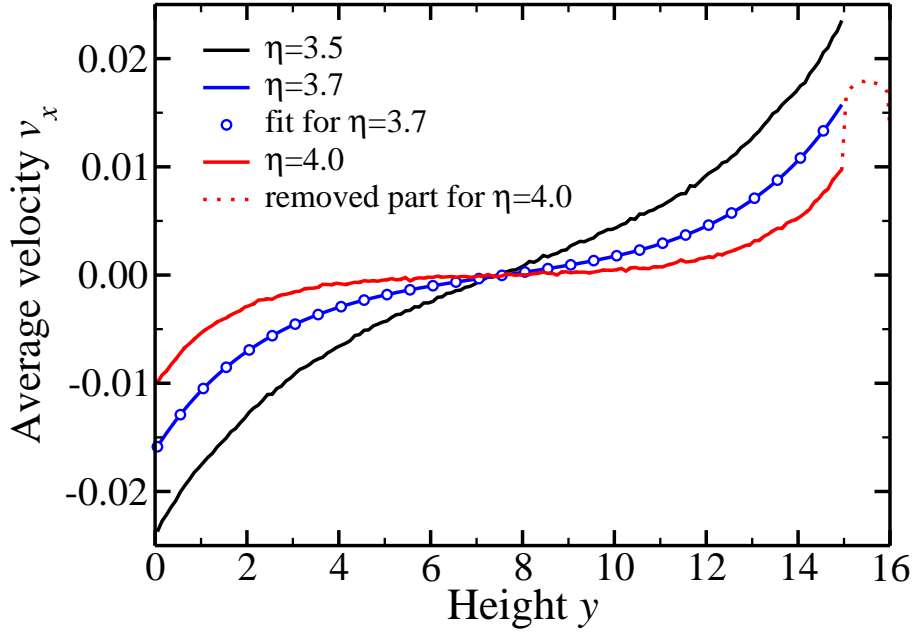}
	\caption{Average velocity in $x$-direction as a function of height $y$ for the RP measurements
	of the metric-free VM for noise values $\eta=3.5$, $3.7$ and $4.0$. Only the lower half of the
	channel is shown. The bottom and top parts of the profiles, {\it i.e.} the velocities inside the
	``feeding''-layers of thickness one were cut off for fitting purposes. This region is shown for
	$\eta=4$ by	the dotted red line, demonstrating the abrupt change of the profile inside the top
	layer of the half-channel. The blue circles show the excellent fit by the function $\propto
	{\rm sinh}(d_1\tilde{y})$ with the shifted height $\tilde{y}=y-L_y/4+0.5$. Simulations have been
	conducted at $L_x=128$, $L_y=32$, $M=5$, $\tau=2$, and $v_0=1$.}
    \label{fig:velocity_profile_dt2_metricfree}
\end{figure}

Figure~\ref{fig:velocity_profile_dt2_metricfree} shows the measured velocity profiles as a function
of height $y$ for the metric-free VM for three different noise values $\eta$ in the disordered phase.
Only the lower half of the channel is shown and the velocities inside the bottom and top layer
were discarded to obtain a better fit. At the border of these ``feeding'' layers the profiles change
abruptly, as shown by the dotted red line for $\eta=4$, which is included in the plot for illustration
but was not used in the fitting procedure. One sees, that the ${\rm sinh}$ function provides a
perfect fit, at least for this set of parameters.

\begin{figure}[htb]
	\includegraphics[width=\figwidth]{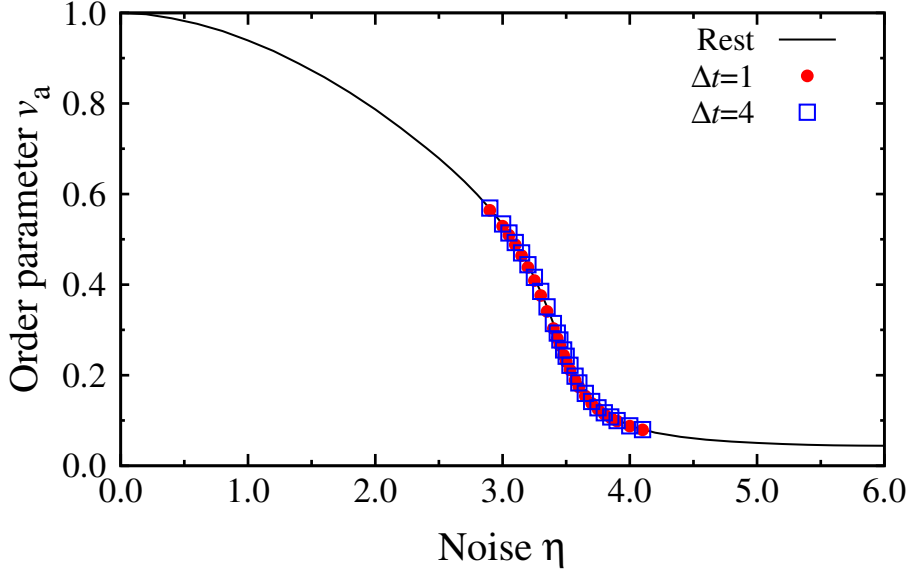}
	\caption{Polar order parameter, $v_{\rm a}$, {\it vs.} noise, $\eta$, for the regular VM at rest
	(solid line) and under shear (symbols). Simulations have been conducted in a $16 \times 16$ box
	with $M=5$, $\tau=2$, $R=1$, and $v_0=1$.}
	\label{fig:order}
\end{figure}

We have also investigated whether shearing the system has an effect on the (average) ordering of the
particle velocities, quantified by the polar order parameter $v_{\rm a}$ [see Eq.~(\ref{eq:va})].
Figure~\ref{fig:order} shows typical results of $v_{\rm a}$ as a function of noise, $\eta$, at rest
and under shear for the regular VM in a $16 \times 16$ box with $M=5$, $\tau=2$, $R=1$, and $v_0=1$.
The data for these three cases are virtually identical, indicating that shear has a negligible
impact on the overall ordering of the particles for the applied shear stresses.

\begin{figure}[htb]
	\includegraphics[width=\figwidth]{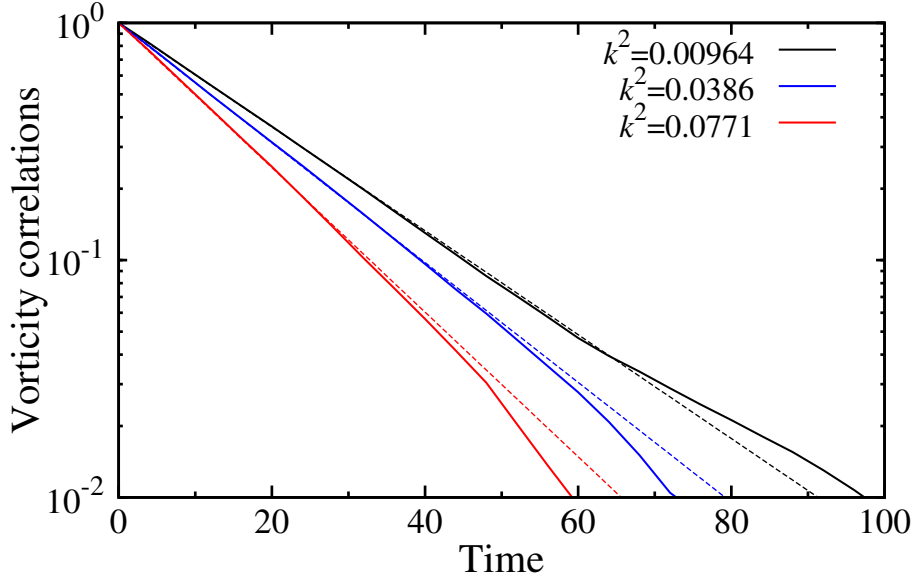}
	\caption{Vorticity correlations, $\la \Omega({\bf k},t)\Omega^*({\bf k},\tilde{t})\ra$, {\it vs.}
	time, $\tilde{t}$, for the metric-free VM at $\eta=3.7$, $M=5$, $\tau=2$, $R_{\rm eff}=1$, and
	$v_0=1$ in a quadratic simulation box with $L_x=L_y=64$. The solid lines show simulation data,
	while the dashed lines correspond to exponential fits.}
	\label{fig:omegaDecay}
\end{figure}

To study the transverse current (or vorticity) correlations, we performed agent-based simulations of
the regular and the metric-free VM with periodic boundary conditions with system sizes $L_x \times
L_y$ ranging from $16 \times 16$ to $64 \times 64$. Neither external forces nor the RP-swapping
procedure were applied. Once a system reached its stationary state, the momentum density was measured
in our simulations through
\begin{equation}
	\label{MIC_DENS_FT}
	\mathbf{\hat{w}}(\mathbf{k},t)=\sum_{j=1}^N \mathbf{v}_j(t) {\rm e}^{i\mathbf{k} \cdot \mathbf{r}_j(t)}, 
\end{equation}
which is the Fourier transformation of the microscopic expression for the momentum density of a
system of $N$ point particles at a particular position $\mathbf{x}$, given by (see for example
Refs. \citenum{forster_75})
\begin{equation}
	\label{MIC_DENS}
	\mathbf{w}(\mathbf{x},t)=\sum_{j=1}^N \mathbf{v}_j(t)\,\delta(\mathbf{x}-\mathbf{r}_j(t)) .
\end{equation}
Here, $\mathbf{v}_j=(v_{j,x},v_{j,y})$ and $\mathbf{r}_j=(r_{j,x},r_{j,y})$ are the velocity and position
of particle $j$, respectively. Inserting the components of the transformed momentum density,
$\mathbf{\hat{w}} = (\hat{w}_x, \hat{w}_y)$, into Eq. (\ref{OMEGA_DEF}), the $z$-component of the
vorticity, denoted by $\Omega$, can be recorded. Typically, the simulations ran for $10^5$ to
$10^6$ iterations after the stationary state has been reached, and $\Omega(\mathbf{k},t)$ was recorded
for a set of small wave vectors.
After completion of the simulations, the stored time series was used to calculate the time-averaged
vorticity correlations, $\la \Omega({\bf k},t)\Omega^*({\bf k},\tilde{t})\ra$. We found that in the
disordered phase, for small $k$ and for not too small time steps, these fluctuations decayed
exponentially, $\propto {\rm exp}(-\mu|t-\tilde{t}|)$, as shown in Fig.~\ref{fig:omegaDecay}.

\begin{figure}[htb]
	\includegraphics[width=\figwidth]{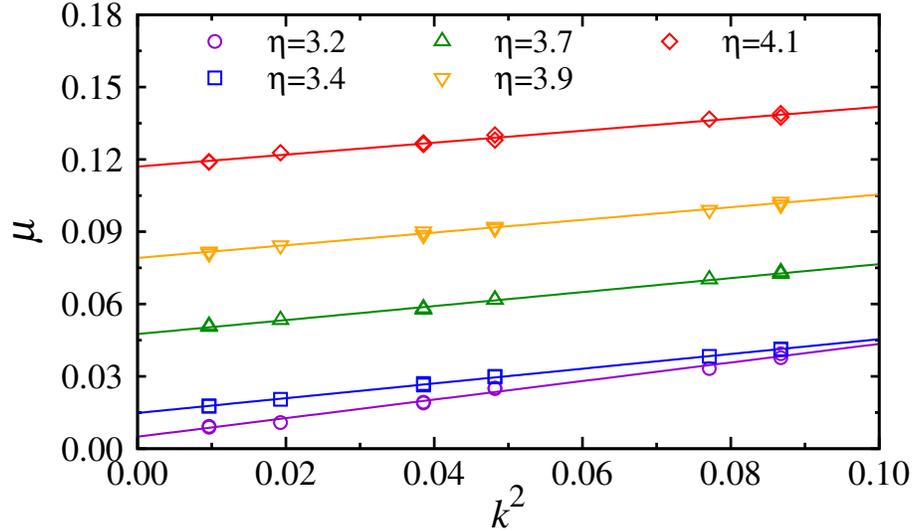}
	\caption{Plot of $\mu$, {\it vs.} $k^2$ for the metric-free VM at $M=5$, $\tau=2$, $R_{\rm eff}=1$,
	and	$v_0=1$ in a quadratic simulation box with $L_x=L_y=64$. Symbols show simulation data, and
	lines show fits according to $\mu=\kappa+\nu k^2$.}
	\label{fig:muFit}
\end{figure}

The decay rate $\mu$ was extracted from exponential fits with different values for the wave vector
$\mathbf{k}$ (see Fig.~\ref{fig:omegaDecay}). We then extracted the transport coefficients $\nu$ and
$\lambda$ from these data by fitting the obtained $\mu$ by the theoretical expectation, 
$\mu = (1-\lambda)/\tau + \nu k^2$ [see Eq.~(\ref{MU_DEF})]. Figure~\ref{fig:muFit} shows exemplary
data for simulations of the metric-free VM at $M=5$, $\tau=2$, $R_{\rm eff}=1$, and $v_0=1$ in a
quadratic simulation box with $L_x=L_y=64$.

\clearpage
\newpage
\subsection{Transport coefficients}
\label{sec:numtrans}
\begin{figure}[htbp]
	\includegraphics[width=\figwidth]{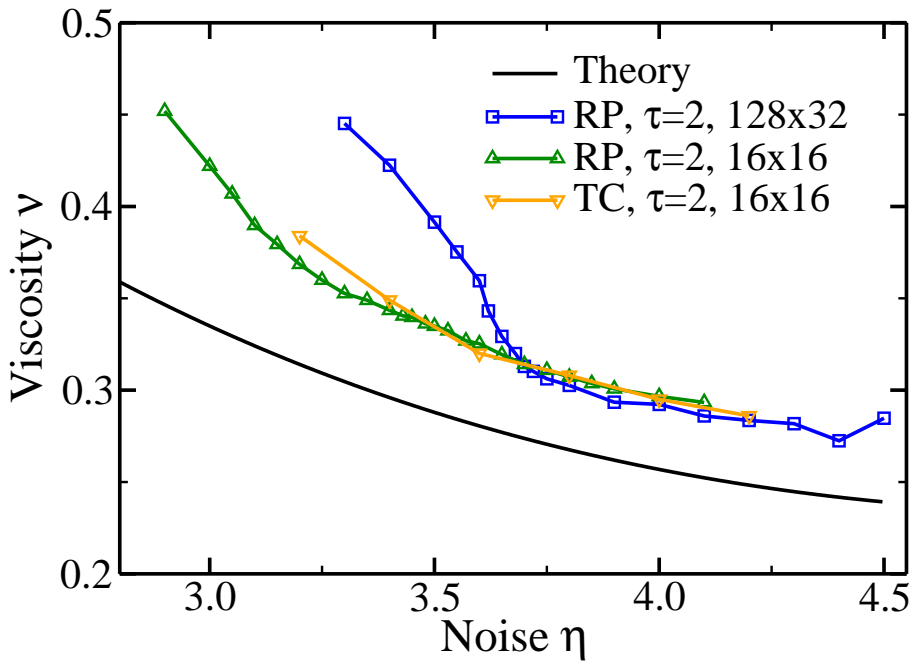}
	\caption{Kinematic viscosity, $\nu$, versus noise, $\eta$, for the regular VM extracted by the 
	RP method for time step $\tau=2$ at two system sizes $L_x=128$, $L_y=32$ (blue squares), and
	$L_x=L_y=16$ (green triangles). The thick black line shows the theoretical prediction by
	Eqs.~(\ref{SUM_VISC}), (\ref{KIN_VISC1}), and (\ref{NU_COLL_ALL}). Results from the TC method
	for $L_x=L_y=16$ are shown by the orange triangles. All simulations conducted for $M=5$, $R=1$,
	and $v_0=1$.}
	\label{fig_viscos_reg_VM1_M5}
\end{figure}

Figure~\ref{fig_viscos_reg_VM1_M5} shows the total shear viscosity, $\nu$, obtained by the RP and
TC methods as a function of noise, $\eta$. The trend of a decreasing viscosity with increasing noise
is the same as in the theoretical prediction. However, the measured values lie consistently by about
$15\%$ to $18\%$ above the mean-field prediction, given by Eqs.~(\ref{SUM_VISC}), (\ref{KIN_VISC1}),
and (\ref{NU_COLL_ALL}), even at parameter ranges where one naively would expect mean-field theory
to hold. This discrepancy is confirmed by the viscosity measurements through the TC method, which
agree rather well with the results from the RP simulations at noise values in the disordered phase. 

\begin{figure}[htb]
	\includegraphics[width=\figwidth]{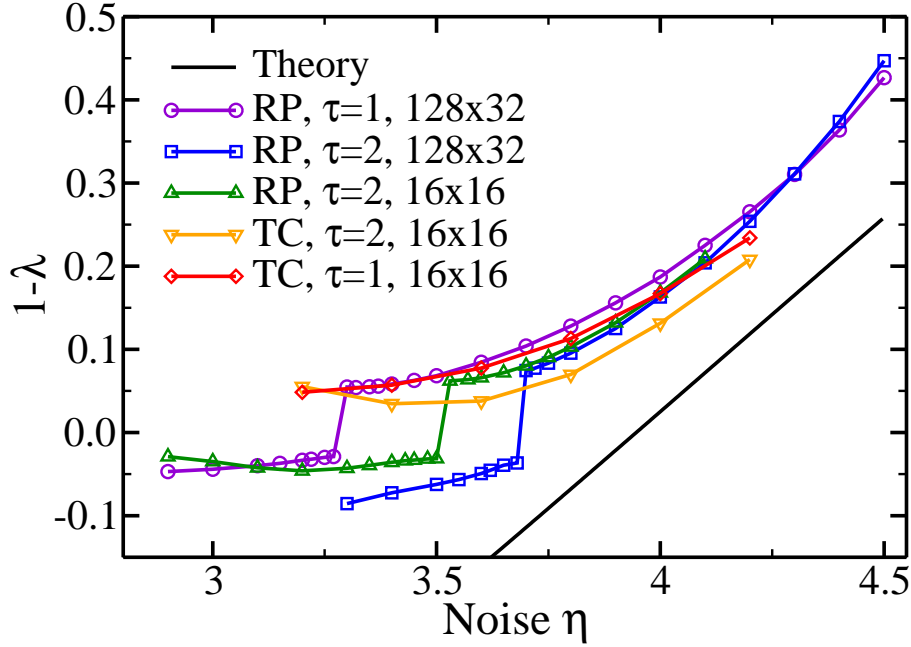}
	\caption{The coefficient $1-\lambda$ {\it vs.} noise, $\eta$, for the regular VM extracted by
	the RP method (purple circles, blue squares and green triangles) for several time steps $\tau$ and
	system sizes $L_x\times L_y$, as indicated. The black line shows the theoretical prediction by
	Eq.~(\ref{DEF_LAM}). Results from the TC method in a small system, $L_x=L_y=16$ are shown for
	$\tau=2$ (orange triangles) and $\tau=1$ (red diamonds). All simulations conducted for $M=5$,
	$R=1$, and $v_0=1$.}
	\label{fig_lambda_reg_VM_M5}
\end{figure}

Figure~\ref{fig_lambda_reg_VM_M5} shows the difference $1-\lambda$ ($\lambda=1$ denotes the threshold
condition for the order/disorder transition) as a function of noise $\eta$ for different system
sizes and time steps. Apart from the values obtained by the RP method, the figure also shows the
mean-field prediction and values obtained by the TC method. Although both the scaled density
$M=\rho\pi R^2=5$ and the ratio between mean free path and interaction radius, $v_0 \tau/R=1$ and
$2$, are rather large, there is a significant deviation between the numerical results and the theory.
This discrepancy indicates that mean-field theory gives quite an inaccurate prediction for the
threshold condition at these parameters. One notices that doubling the time step, $\tau$, reduces
the deviation to mean-field theory only slightly.  

\begin{figure}
	\includegraphics[width=\figwidth]{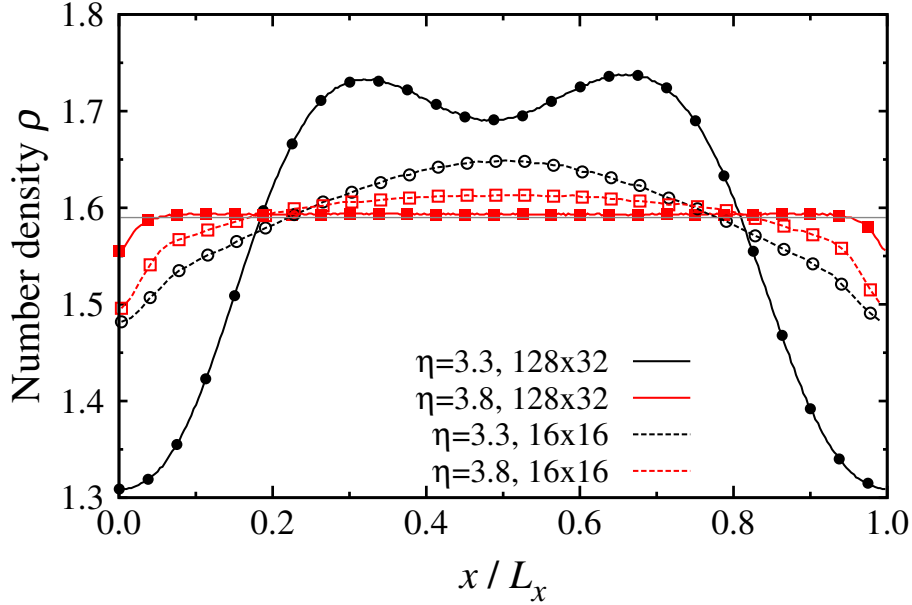}
	\caption{Density distribution along the normalized $x$-axis, taken relative to the system's
	center of mass. Solid lines with filled symbols show the results for the $128 \times 32$ systems,
	whereas dashed lines with open symbols show the results for the $16 \times 16$ system. The grey
	horizontal line indicates the average density in the system. The symmetric double peak of the
	density distribution in the $128 \times 32$ systems for $\eta=3.3$ originates from averaging
	multiple density waves, traveling in both the $+x$ and $-x$ directions over the course of the
	simulation.}
	\label{fig:densityWaves}
\end{figure}

Furthermore, the parameter $\lambda$ seems to jump from a positive value to a negative one around
some critical value if the noise is decreased. This behavior could be due to the appearance of
density waves right at the onset of collective motion, that render the order/disorder transition
discontinuous. A quantitative, mean-field theory of this mechanism in the regular VM is presented
in Ref. \citenum{ihle_13}. However, density waves only occur in sufficiently large systems; Figure
\ref{fig:densityWaves} shows the density distribution of particles along the $x$-direction, taken
relative to the center of mass of the system for the large $128 \times 32$ systems as well as the
small $16 \times 16$ systems at $\tau=2$. Measurements have been taken for noise values in the
ordered ($\eta = 3.3 < \eta_{\rm c}$) and disordered regime ($\eta = 3.8 > \eta_{\rm c}$). The
large systems exhibit distinct density waves when $\eta < \eta_{\rm c}$, whereas the small systems
did not develop any such density waves. Note that applying shear did not have any appreciable effect
on the formation of density waves, but the wave fronts appear to be less sharp due to the emerging
${\rm sinh}$-shaped velocity profile (see Fig.~\ref{fig:velocity_profile_dt2_metricfree}).

\begin{figure}[htb]
	\includegraphics[width=\figwidth]{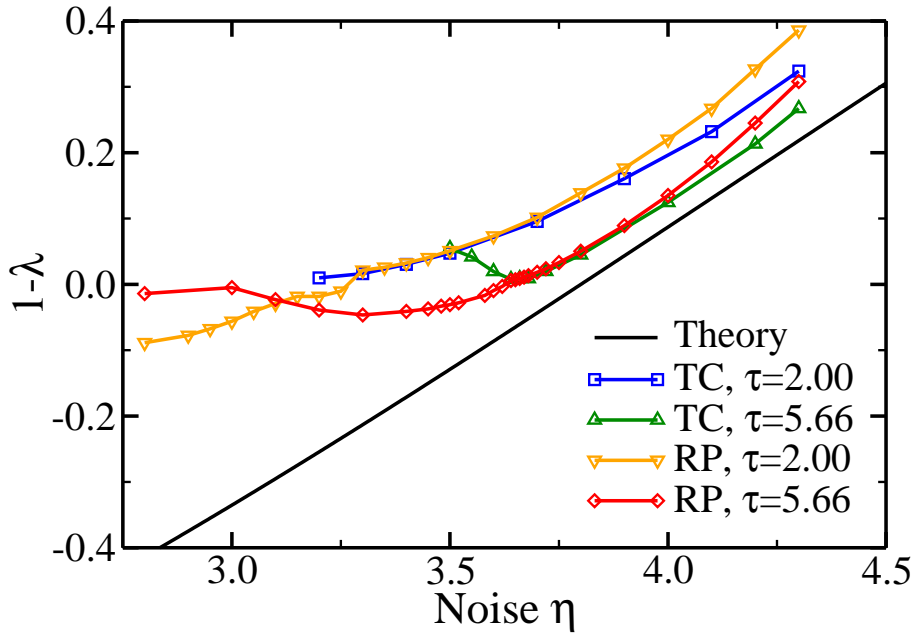}
	\caption{The coefficient $1-\lambda$ {\it vs.} noise, $\eta$, for the metric-free VM, extracted
	by the RP and TC method for time steps $\tau=2.0$ and $\tau=5.66$, as indicated. The black line
	shows the theoretical prediction from Ref. \citenum{chou_12}. Simulations have been conducted at
	$M=5$, $R_{\rm eff}=1$, and $v_0=1$ in a quadratic simulation box with $L_x=L_y=64$	for the TC
	runs and in a box with $L_x=128$, $L_y=32$ for the RP simulations. The error bars on the TC
	data are smaller than the symbols.}	
	\label{fig_lambda_metricfree_trans}
\end{figure}

Nevertheless, even in small systems there might be strong density fluctuations and/or transient
clusters at the threshold to collective motion that are precursors of the discontinuous phase
transition which is observed in larger systems. In order to test the hypothesis that the jump in the
measured values of $1-\lambda$ is caused by those density fluctuations and are not artifacts of the
RP method, we also performed measurements for the VM with metric-free interactions. As shown in
Fig.~\ref{fig_lambda_metricfree_trans} and comparing with Fig.~\ref{fig_lambda_reg_VM_M5}, these
jumps are about a factor of three smaller in the metric-free model and are hardly noticeable in the
plot. Further note, that mean-field theory underestimates the coefficient $1-\lambda$ also for the
metric-free model. However, increasing the time step from $\tau=2$ to $\tau=5.66$ leads to better
agreement with mean-field theory, as expected. In general, it appears that the agreement with
mean-field theory is better for the metric-free model than for the regular VM.

Still, we cannot completely rule out that the observed jump in $1-\lambda$ might be an
artifact of the assumed hydrodynamic theory which we used to evaluate the RP measurements. In that
case, the jump could be interpreted as an error bar in the determination of the momentum gain
coefficient $\lambda$ or, alternatively, could simply mean that the RP method is not very reliable
in the ordered phase. These interpretations are consistent with an alternative measurement of
$\lambda$ by the TC method. In particular, comparing the green triangles to the orange triangles in
Fig.~\ref{fig_lambda_reg_VM_M5}, which correspond to the same set of parameters but different
methods, we see good agreement in the disordered phase with a difference between the two curves that
is smaller than the jump in the green curve.

\begin{figure}[htb]
	\includegraphics[width=\figwidth]{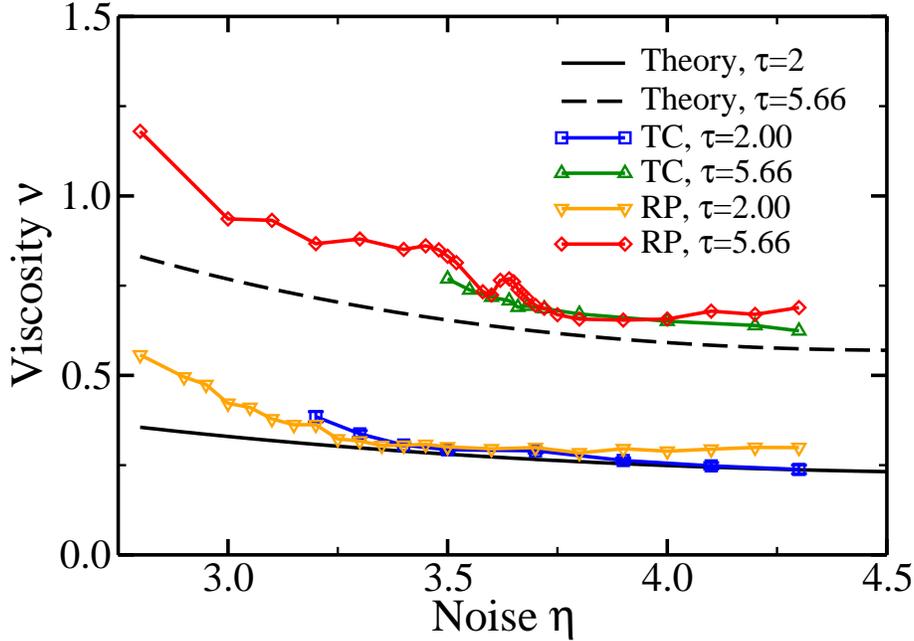}
	\caption{Kinematic viscosity, $\nu$, {\it vs.} noise, $\eta$, for the metric-free VM extracted
	by the RP and TC method for time steps $\tau=2.0$ and $\tau=5.66$. The solid black line shows
	the theoretical result for $\tau=2$, whereas the dashed black curve is the theoretical
	prediction with	$\tau=5.66$ from Ref. \citenum{ihle_19}. Simulations have been conducted at
	$M=5$, $R_{\rm eff}=1$,	and $v_0=1$ in a quadratic simulation box with $L_x=L_y=64$ for the	TC
	runs and in a box with $L_x=128$, $L_y=32$ for the RP simulations. The error bars on the TC data
	are equal to or smaller than the symbols.}	
	\label{fig_viscos_large_dt_metricfree}
\end{figure}

Similar agreement between the RP and TC method is seen in Fig.~\ref{fig_viscos_large_dt_metricfree},
which shows measurements of the viscosity for the metric-free model (see Sec.~\ref{sec:Vicsek} for a
definition of the model). Although near the threshold noise, $\eta_{\rm c}$, excellent agreement
between the TC and the RP method occurs, deviations are observed at larger noise. At these larger
noises, the TC method appears to be more accurate than the RP method or at least seems to require
less fine-tuning of numerical parameters such as the appropriate thickness of the feeding layers,
swap times $\Delta t$ and so on. 

To quantify the errors of the RP method we fitted three different sections of the velocity profiles
for the parameters of Figs.~\ref{fig_lambda_metricfree_trans} and \ref{fig_viscos_large_dt_metricfree}
at $\eta=4.0$ and for both $\tau=2$ and $\tau=5.66$. This leads to different fitting coefficients
$d_0$ and $d_1$, which consequently lead to different predictions when plugged into Eqs.~(\ref{NU_FIND}),
(\ref{LAM_FIND}) or (\ref{LAM_FIND_ORDER}). For the runs with $\tau=2$ we found rather small errors,
about $5\%$ for $\nu$ and $2\%$ for $\lambda$. However, for the larger time step $\tau=5.66$
the errors become huge if sections of the profile are picked for fitting that either only include
profile parts from near the center of the sample or only parts from the vicinity of the feeding
layers. In this worst case scenario, one obtains a $100\%$ error in both $\nu$ and $\lambda$.
However, by comparing the green with the red curve in Fig.~\ref{fig_lambda_metricfree_trans}, 
the error for $\lambda$ actually appears to be only around $15\%$ to $18\%$ at the largest noises
and even smaller for $\nu$. Nevertheless, this serves as a warning that the thickness of the feeding
layer and the channel height $L_y$ must be chosen carefully and large enough compared to the
mean-free path.

Let us now focus on the effect of time step for otherwise identical conditions (purple circles and
blue squares in Fig.~\ref{fig_lambda_reg_VM_M5}). Here, we observe that the jump, which indicates
the order/disorder transition, occurs at a smaller noise value in the system with the smaller time
step $\tau=1$. This shift is a well-known effect in the standard VM, which has been reported for
instance in Refs.~\citenum{chate_08, ihle_11}. It has been shown that at large mean free paths,
corresponding to large time steps, the threshold noise converges to the mean field prediction
\cite{ihle_11}, which follows from Eq. (\ref{DEF_LAM}) by setting $\lambda$ equal to one. This
mean-field prediction for $\eta_{\rm c}$ does not depend on particle velocity, time step or
interaction radius. However, as observed in Ref. \citenum{chate_08}, the actual value of $\eta_{\rm c}$
becomes smaller by up to a factor between two and three if the mean free path (or $\tau$ in our
case) is reduced. This deviation from mean-field theory is attributed to correlation effects which
grow at decreasing mean free path. Although a ring-kinetic theory for correlation effects in the
standard VM was attempted in Ref.~\citenum{chou_15}, it fell short of explaining the dependence of
$\eta_{\rm c}$ on the mean free path.

A similar dependence of the threshold noise on the mean free path can be seen in
Fig.~\ref{fig_lambda_metricfree_trans} which shows $1-\lambda$ for the metric-free model at two
different mean free paths, $v_0\tau=2$ and $5.66$. Here, the threshold noises are $\eta_{\rm c}
\approx 3.24$ and $\eta_{\rm c} \approx 3.63$, respectively. Around these noise values one observes
a tiny jump of $1-\lambda$ in the figure. Additionally, one sees that the RP method becomes less
accurate deep in the ordered phase but also further in the disordered phase, away from the threshold,
see also the orange curve in Fig.~\ref{fig_viscos_large_dt_metricfree} for $\tau=2$ and $\eta \geq 3.9$. 
This is partly because if $\lambda$ differs significantly from one, the velocity profile decays
rapidly towards the middle of each half-channel, making a fit by a $\sinh$-function less reliable.
Furthermore, at the larger mean free path, $v_0\tau=5.66$ and a ``feeding layer'' smaller than this
length (as used in our simulations), the discreteness of the dynamics impacts the velocity profile
which deviates from  a $\sinh$-function. Fitting it anyway by such a function creates a rather large
error. This is especially visible in Fig.~\ref{fig_lambda_metricfree_trans} at noises $\eta \geq 4.0$ 
and large $\tau=5.66$.

We noticed a significant $k$-dependence of the viscosity near the order/disorder threshold at certain
parameter values. In particular, fitting the decay constant $\mu$ by a function $\kappa+\tilde{\nu}
k^{\beta}$ with three free parameters sometimes led to an exponent $\beta$ smaller than two, at least
in the range of $k$-values we investigated. We checked that $\beta$ approaches the value two at
larger mean free paths and further away from the threshold, {\it i.e.} at larger noise. Since this
effect seems to be more pronounced than in regular fluids near criticality, we think that it is a
result of the velocity alignment interaction. A detailed numerical and analytical investigation of
this behavior is beyond the scope of the paper but is subject of current research.

\begin{figure}[htb]
	\includegraphics[width=\figwidth]{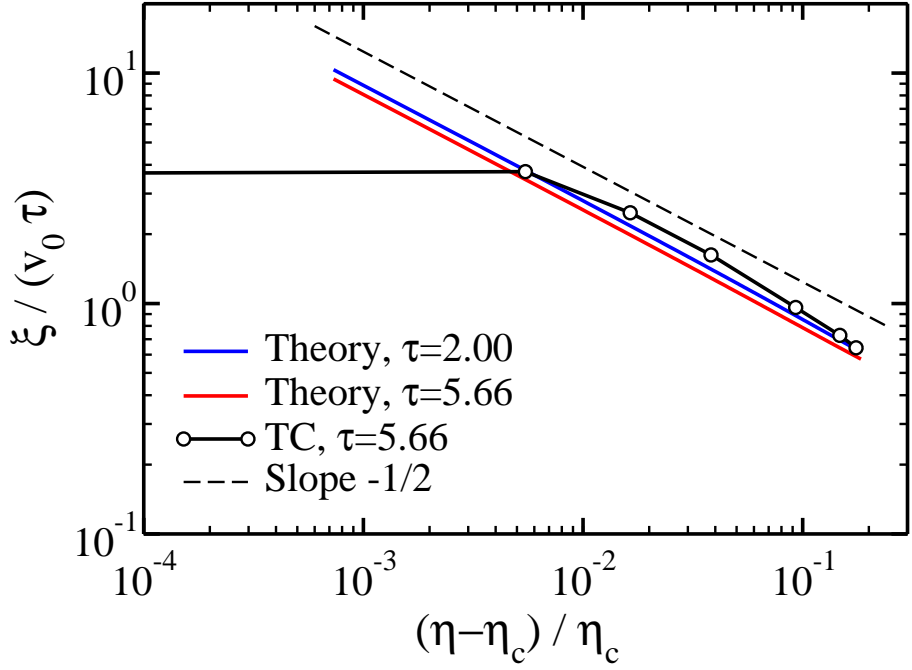}
	\caption{Mean-field correlation length $\xi$ in units of the mean free path $\Lambda=v_0\tau$
	{\it vs.} relative distance to the threshold noise, $(\eta-\eta_{\rm c})/\eta_{\rm c}$, for the
	metric-free VM. Instead of measuring the actual correlation length directly, $1-\lambda$ and the
	viscosity $\nu$	were obtained from measured TC and then inserted in the formula
	$\xi=\sqrt{\nu \tau/(1-\lambda)}$ (open circles). The blue and red lines show the theoretical
	prediction for time steps $\tau=2$ and $5.66$, respectively. The dashed line represents a power
	law decay with exponent $-1/2$.	Simulations have been conducted at $M=5$, $R_{\rm eff}=1$, and
	$v_0=1$ in a quadratic simulation box with $L_x=L_y=64$.}
	\label{fig_correlation_length1}
\end{figure}

The quantity, $\xi \equiv \sqrt{\nu/\kappa}$ has units of length, and within kinetic theory, it can
be established \cite{ihle_19} as a mean-field approximation to the correlation length. At a continuous
phase transition, the correlation length should diverge with the critical exponent $\bar{\nu}$ as
\begin{equation}
	\label{CRIT_EXP1}
	\xi \sim (\eta-\eta_{\rm c})^{-\bar{\nu}}
\end{equation}
with the usual mean-field exponent $\bar{\nu}=1/2$, see for example Ref.~\citenum{binney_92}.
Inserting our measured values of $\kappa$ and $\nu$ into the expression for $\xi$, and plotting this
as a function of the relative distance to the threshold noise, $(\eta-\eta_{\rm c})/\eta_{\rm c}$,
the divergence with the mean-field exponent of $1/2$ is reproduced rather well, as shown in
Fig.~\ref{fig_correlation_length1}. Of course, this result does not rule out that the actual
correlation length diverges with an exponent different from $1/2$. Measuring of actual critical
exponents requires careful finite size scaling and is beyond the scope of this paper.

\subsection{Evaluation of Green-Kubo relations}
GK relations \cite{green_54, kubo_57, zwanzig_65, forster_75} provide a convenient way to measure
transport coefficients in equilibrium Molecular Dynamics or other particle-based simulation methods
of regular fluids. Typically, these relations are used to obtain the self-diffusion coefficient and
the viscosity. The kinetic part of the viscosity describes the convection of the transverse momentum
by a particle. More specifically, every particle that moves in the $x$-direction with some velocity
$v_x$ carries its transverse momentum $p_y=mv_y$ with it; when it eventually collides with another
particle, it has transferred $y$-momentum in the $x$-direction. This mechanism leads to the
appearance of the off-diagonal element $\sigma_{xy}^{\rm kin}(t)=m\sum_{j=1}^N v_{j,x}(t)v_{j,y}(t)$
of the kinetic stress tensor in the derivation of the corresponding GK relation by the
projector-operator method for regular fluids. A similar derivation for the MPCD-fluid can be found
in Refs. \citenum{ihle_03a, ihle_03b}.

It seems plausible that the same mechanism of momentum transport acts also in generalized fluids,
such as the VM, that neither respect momentum conservation nor detailed balance. Indeed, it was shown
\cite{ihle_16} that the analytical evaluation of the usual GK relation for the kinetic part of the
viscosity
\begin{equation}
\label{GREEN_KUB1}
	\nu_{\rm kin}={\tau\over N k_{\rm B} T}\left[ {1\over 2} \la \sigma_{\rm kin}^2(0)\ra +
	{\sum_{n=1}^{\infty}} \la \sigma_{\rm kin}(n\tau)\sigma_{\rm kin}(0)\ra \right]
\end{equation}
within the mean-field assumption of molecular chaos and setting the temperature $k_B T$ equal to
$mv_0^2/2$, leads to an expression for $\nu_{\rm kin}$ which is identical to the one obtained from
the Chapman-Enskog theory of the VM, Eq. (\ref{KIN_VISC1}). Thus, at least at the mean-field level,
the validity of the GK relation, Eq.~(\ref{GREEN_KUB1}), has been proven. Although a microscopic
derivation of the GK relation for the VM has not been performed yet, its correctness beyond
mean-field seems likely, and we use it here anyway.
\begin{figure}[htb]
	\includegraphics[width=\figwidth]{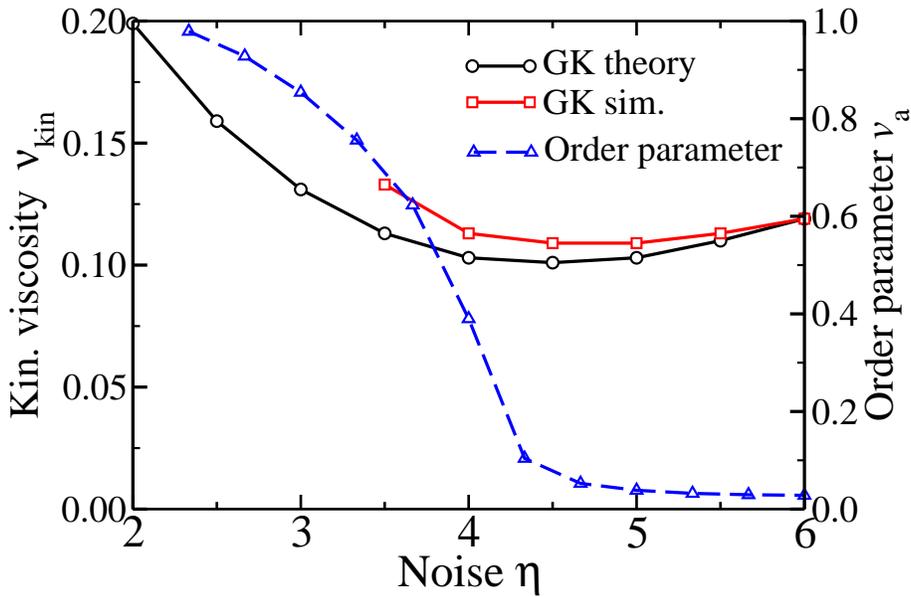}
	\caption{Kinetic part of the viscosity, $\nu_{\rm kin}$, {\it vs.} noise, $\eta$, for the regular
	VM. Red squares show measurements in the disordered state by means of Eq. (\ref{GREEN_KUB1}),
	while black circles correspond to the prediction by mean-field kinetic theory, Eq. (\ref{KIN_VISC1}).
	The blue dashed line is the measured polar order parameter (right axis). Simulations have been
	conducted for $M=5$, $\tau=1$, $v_0=1$, and $R=1$.}
	\label{green-kubo1}
\end{figure}

\begin{figure}[htb]
	\includegraphics[width=\figwidth]{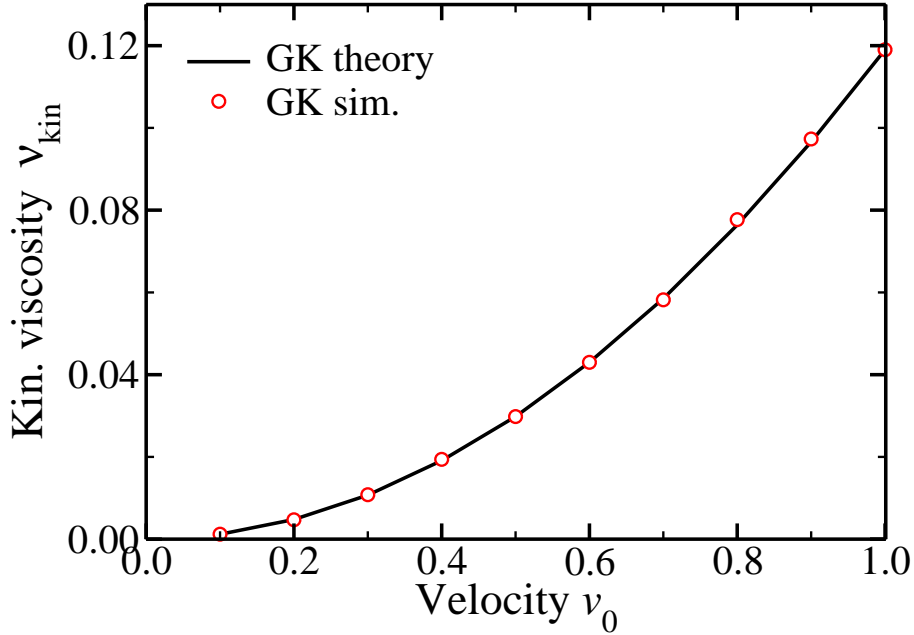}
	\caption{Kinetic part of the viscosity, $\nu_{\rm kin}$, {\it vs.} velocity, $v_0$, for the regular
	VM at very large $\eta=6$. Red circles show measurements in the disordered state by means of Eq. 
	(\ref{GREEN_KUB1}), while the solid black line corresponds to the evaluation of this expression by
	mean-field kinetic theory. Simulations have been conducted at $M=5$, $\tau=1$, and $R=1$.}
	\label{green-kubo2}
\end{figure}

Figure~\ref{green-kubo1} shows the kinetic viscosities measured in direct simulations of the VM
without shear gradient by means of Eq.~(\ref{GREEN_KUB1}) as a function of noise, $\eta$, in
comparison with the theoretical (mean-field) expression. For higher noise, where the theory should
become more accurate, we find excellent agreement for a large range of particle velocities, $v_0$,
as shown in Fig.~\ref{green-kubo2}. However, even at the threshold to ordered motion, the deviation
is only around $15\%$. These results support the validity of the mean-field derivation of the
analytical expression for $\nu_{\rm kin}$, Eq. (\ref{KIN_VISC1}), which was first reported in
Ref.~\citenum{ihle_11}. Understanding and reducing the deviations between kinetic theory and
agent-based simulations will be left for future studies.

\section{Conclusions}
\label{sec:sum}
By now, one can find many derivations of hydrodynamic equations from the microscopic interactions of
active particle systems in the literature. These derivations are often complicated and involve
several more or less severe approximations. Therefore, the validity of the obtained expressions is
not {\it a priori} clear, and it would be useful to verify them. In this manuscript, we developed
a hydrodynamic theory for both the standard and the metric-free version of the Vicsek model (VM) of
self-propelled particles, when shear is applied through M{\"u}ller-Plathe's reverse perturbation
(RP) method. Feeding momentum into the boundaries of a channel filled with self-propelled particles
led to an almost exponential decay of the flow speed towards the center of the channel due to the
lack of momentum conservation. We demonstrated how fitting this decay with an analytical solution of
the hydrodynamic equations for the VM allows extracting the two transport coefficients, namely the
shear viscosity, $\nu$, and the momentum amplification coefficient, $\lambda$. In order to compare
with existing kinetic theories, an improvement of a previous derivation of the viscosity from an
Enskog-like kinetic theory was required. This calculation resulted in a new explicit formula for
the missing contribution -- the collisional part of the viscosity. For a typical choice of
parameters from Vicsek's original paper, we showed that this collisional contribution is larger by
a factor of $\approx 10^4$ than the previous prediction for the viscosity.

To verify our theory, we performed agent-based simulations of both the standard and the metric-free
version of the VM. We measured the transport coefficients $\nu$ and $\lambda$ using two different
methods, namely the RP method and the transverse current fluctuation method (TC). In the disordered
phase and not too far from the threshold to collective motion, excellent agreement between the
measurements of $\nu$ was found. These findings verify our extension of M{\"u}ller-Plathe's RP
method to active particle systems. 

Further, we found reasonable agreement when comparing our measurements of $\nu$ with the predictions
from mean-field kinetic theory. However, the measured viscosities were consistently higher by $15\%$
to $18\%$ than the predicted ones. To elucidate the origin of this systematic discrepancy, we also
measured the kinetic part of the shear viscosity, $\nu_{\rm kin}$, using the Green-Kubo (GK)
approach. In the GK calculations we observed very good agreement between theory and measurements at
large noise, close to the maximum noise of $\eta_{\rm max}=2\pi$. However, close to the threshold
to collective motion we found a similar difference of about $15\%$ as in the RP and TC measurements.
Because most of our measurements were done at rather large time step $\tau \geq 1$, where the
viscosity is dominated by its kinetic part, we hypothesize that most of the discrepancy in $\nu$
between theory and simulation is due to the invalidity of the mean-field assumption in the analytical
calculation of $\nu_{\rm kin}$. Therefore, future theoretical efforts to improve the expression for
the viscosity should focus on this contribution.

The agreement between the results for $\lambda$ obtained by the RP and TC method was also very good
but still not as good than for the viscosities. Moreover, we observed that the mean-field prediction
for $\lambda$ only became accurate at very large mean-free paths, where there is large mixing of
particles and where the assumption of molecular chaos should become valid.

While our results support the correctness of the novel mean-field calculation of the collisional
part of the viscosity as well as the validity of earlier results from an Enskog-like kinetic theory,
they however underline previous concerns about mean-field assumptions and the relevance of correlation
effects in active matter systems. It appears that even at a large average number of interaction
partners, $M=5$, and a mean free path that is twice as large as the interaction range, pre-collisional
correlations still significantly influence the transport coefficients.

\section{Acknowledgments}
A.N. acknowledges funding from the German Research Foundation (DFG) under the project number
NI 1487/2-1. Computing time was granted on the supercomputer Mogon at Johannes Gutenberg University
Mainz (www.hpc.uni-mainz.de).

\appendix
\section{Kinetic theory for the Vicsek model}
\label{sec:appendixA}
\subsection{Introduction to Enskog-like kinetic theory}
\label{sec:enskog}
In the VM, a given particle $i$ is described by its location ${\bf x}_i$ and the angle $\theta_i$ of
its velocity vector. Hence, the microstate of a system of $N$ particles corresponds to a point in
$3N$-dimensional phase space. The time-evolution of the VM in this phase space is Markovian, since
information about microstates from earlier times is irrelevant for further evolution. Hence, we can
write down an exact evolution equation for the $N$-particle probability density $P$ of the
corresponding Markov chain,
\begin{equation}
	P({\bf B},t+\tau)=\int P({\bf A},t)\;W_{AB}\; {\rm d}{\bf A}\,,
	\label{MASTER1}
\end{equation}
which describes the transition from microscopic state ${\bf A}$ to state ${\bf B}$ during one
time step with transition probability $W_{AB}$. The state of the system at time $t+\tau$ is given by
the vector, ${\bf B} \equiv (\theta^{(N)}, {\bf X}^{(N)})$, where $\theta^{(N)} \equiv (\theta_1,
\theta_2, \ldots, \theta_N)$ contains the flying directions of all $N$ particles, and ${\bf X}^{(N)}
\equiv ({\bf x}_1, {\bf x}_2,\ldots, {\bf x}_N)$ describes all particle positions. The initial
microscopic state at time $t$ is denoted as ${\bf A}\equiv(\tilde{\theta}^{(N)},
{\bf \tilde{X}}^{(N)})$. The integral over the initial state translates to $\int\,{\rm d}{\bf A}
\equiv \prod_{i=1}^N\int_{-\pi}^{\pi}\,{\rm d}\tilde{\theta}_i\int\,{\rm d}{\bf \tilde{x}}_i$, where
pre-collisional angles and positions are given by $\tilde{\theta}_i$ and ${\bf \tilde{x}}_i$, respectively.
The transition probability $W_{AB}$ encodes the microscopic collision rules,
\begin{equation}
	W_{AB}=\prod_{i=1}^N \delta({\bf \tilde{x}}_i-{\bf x}_i+\tau {\bf v}_i)
	\,\int_{-\pi}^{\pi}\, w_n(\xi_i)\;\,\hat{\delta}(\theta_i-\xi_i-\Phi_i)\;\,{\rm d}\xi_i\,,
	\label{TRANS1}
\end{equation}
and consists of two parts: the first $\delta$-function describes the streaming step which changes
particle positions. The second part contains the periodically continued delta function,
$\hat{\delta}(x)=\sum_{m=-\infty}^{\infty}\delta(x+2\pi m)$, which accounts for the modification of
angles in the collision step. The particle velocities ${\bf V}^{(N)} \equiv ({\bf v}_1,{\bf v}_2,\ldots,
{\bf v}_N)$, are given in terms of angular variables $\theta_i$,
\begin{equation}
	{\bf v}_i=(e_x, e_y)=v_0\,(\cos{\theta_i}, \sin{\theta_i})\,.
	\label{DEF_VELO}
\end{equation}
For the standard VM, the noise distribution $w_n$ is given by
\begin{equation}
\label{ETADEF}
 w_n(\xi) = \begin{dcases*}
        {1\over \eta}     & for $-{\eta\over 2}\le \xi \le {\eta\over 2}$\\
        0                 & elsewhere.
        \end{dcases*}
\end{equation}
with noise strength $\eta$. Solving Eq.~(\ref{MASTER1}) is intractable without major simplification.
The common way to proceed is to use Boltzmann's molecular chaos approximation by assuming that the
particles are uncorrelated just prior to every microscopic interaction \cite{foot1}. This approximation amounts
to a factorization of the $N$-particle probability into a product of one-particle probabilities,
{\it i.e.} $P(\theta^{(N)}, {\bf X}^{(N)}) = \prod_{i=1}^N P_1(\theta_i, {\bf x}_i)$ on the right
hand side of Eq.~(\ref{MASTER1}). Because molecular chaos neglects pre-collisional correlations, the
resulting theory has a mean-field nature. By integrating out all particles except one -- the so-called
focal particle -- in Eq.~(\ref{MASTER1}), an Enskog-like equation for the distribution function
$f=N P_1$ is obtained,
\begin{equation}
	f({\bf x} + \tau {\bf v}, \theta, t+\tau) = C \circ	f({\bf x}, \theta, t))\,,
	\label{ENSKOG_MAIN}
\end{equation}
where $C$ is an Enskog collision operator for multi-particle collisions. In the thermodynamic limit,
$N \to \infty$, $L \to \infty$, and $\rho_0=N\L^2=const.$, this operator is given by
\begin{eqnarray}
\nonumber
C\circ f({\bf x},\theta, t)&=&
{1\over \eta}
\int_{-\eta/2}^{\eta/2}
{\rm d}\xi
\bigg\langle \bigg\langle
\sum_{n=1}^{\infty}
{{\rm e}^{-M}\over n!}
\,n\, \\
\label{ENSKOG1}
& &\times f({\bf x}, \tilde{\theta}_1, t)
\,\hat{\delta}(\theta-\xi-\Phi_1)
\,\prod_{i=2}^n f({\bf x}_i, \tilde{\theta}_i,t)
\bigg\rangle_{\tilde{\theta}} \bigg\rangle_{\bf x}
\,.
\end{eqnarray}
Here, $\langle \ldots \rangle_{\bf x} = \int_{\odot} \ldots \,{\rm d}{\bf x}_2\,{\rm d}{\bf x}_3
\ldots {\rm d}{\bf x}_n$ denotes the integration over all positions of the particles $2, 3, \ldots n$
inside the collision circle, and $\langle \ldots \rangle_{\tilde{\theta}}=\int_0^{2\pi} \ldots
{\rm d}\tilde{\theta}_1\, {\rm d}\tilde{\theta}_2 \ldots {\rm d}\tilde{\theta}_n$ refers to the
integration over the pre-collisional angles of all $n$ particles inside the circle. The average
angle of the focal particle $i=1$, $\Phi_1$ is defined in Eq.~(\ref{COLLIS}) and is a function of
both the pre-collisional angles and the positions of all particles. For more details on the
derivation of Eq.~(\ref{ENSKOG_MAIN}) and a discussion of the molecular chaos assumption, see
Refs.~\citenum{ihle_14_a} and \citenum{ihle_16}.

\subsection{Calculation of the collisional viscosity $\nu_{\rm coll}$}
\label{sec:calc}
To calculate the collisional viscosity, we will heavily rely on the notations and equations presented
in Ref. \citenum{ihle_16}, which are too lengthy to be repeated here in full detail \cite{foot2}.
There, a Chapman-Enskog expansion (CE) \cite{cercignani_88, enskog_21, chapman_52}, which is basically
an elaborated gradient expansion, was constructed to obtain hydrodynamic equations of the VM. To
systematically account for gradients in the hydrodynamic fields, a dimensionless ordering parameter
$\epsilon$ had been introduced, which was set to unity at the end of the calculation. As a
``byproduct'' of the CE, expressions for the transport coefficients and the equation of state in
terms of microscopic parameters were obtained.

The non-standard CE procedure of Ref.~\citenum{ihle_16} starts with a Taylor expansion of the left
hand side of Eq.~(\ref{ENSKOG_MAIN}), in which spatial gradients are scaled as $\partial_{\alpha}
\to \epsilon\partial_{\alpha}$, and multiple time scales $t_i$, whose physical meaning is explained
at the end of this Appendix, are introduced in the temporal gradients, 
\begin{equation}
	\partial_t \equiv
	\partial_{t_0} + \epsilon \partial_{t_1} + \epsilon^2 \partial_{t_2} + \epsilon^3 \partial_{t_3}\,.
	\label{GRAD_EXPAND}
\end{equation}
In addition, the distribution function $f$ and the collision integral, {\it e.g.} the right hand side
of Eq.~(\ref{ENSKOG_MAIN}), are expanded in powers of $\epsilon$,
\begin{eqnarray}
	\nonumber
	f &=& f_0+\epsilon f_1+\epsilon^2 f_2+\epsilon^3 f_3 \\
	\label{EXPAND_F}
	C \circ f &=& C_0+\epsilon C_1+\epsilon^2 C_2+\epsilon^3 C_3
\end{eqnarray}
In Ref.~\citenum{ihle_16} it was shown that the expansion of the distribution function $f$ in Eq.
(\ref{EXPAND_F}) can be identified as an angular Fourier series,
\begin{eqnarray}
	\label{FOURIERB1}
	f_0({\bf x},t) &=& {\rho({\bf x},t) \over 2\pi} \\
	\label{FOURIERC1}
	f_n({\bf x},\theta,t)&=&{1\over \pi v_0^n}\left[a_n({\bf x},t)\cos{(n\theta)}+b_n({\bf x},t)\sin{(n\theta)}\right]
\;\;{\rm for}\;n>0
\end{eqnarray}
with Fourier coefficients $a_i$ and $b_i$. Thus, the reference state $f_0$ of the CE, that is, the
leading order contribution to $f$, coincides with the zero mode of the Fourier series.

To obtain Toner-Tu-like equations, the CE expansion has to be performed up to third order in
$\epsilon$, given the chosen scaling of Eqs.~(\ref{GRAD_EXPAND}) and (\ref{EXPAND_F}). Collecting
terms in orders of $\epsilon$ leads to a hierarchy of coupled equations for the temporal evolution
of $f_i$, which are given by Eqs.~(22-25) in Ref.~\citenum{ihle_16}. These equations contain the
microscopic velocity vector, given in Eq.~(\ref{DEF_VELO}).

The goal is to obtain macroscopic equations for the first two moments of $f$, namely the particle
density $\rho$ and the momentum density vector ${\bf w}=(w_x,w_y)$, which are the ``slow'' fields in
this problem,
\begin{eqnarray}
	\nonumber
	\rho&=&\int_0^{2\pi}f\,{\rm d}\theta \\
	\nonumber
	w_x&=&\rho u_x=\int_0^{2\pi}e_x\, f\,{\rm d}\theta=\int_0^{2\pi} v_0 \cos{\theta}\, f\,{\rm d}\theta \\
	\label{DEF_HD_FIELDS}
	w_y&=&\rho u_y=\int_0^{2\pi}e_y\, f\,{\rm d}\theta=\int_0^{2\pi} v_0 \sin{\theta}\, f\,{\rm d}\theta\,. 
\end{eqnarray}
where ${\bf u}=(u_x,u_y)={\bf w}/\rho$ denotes the macroscopic flow velocity. To proceed, velocity
moments of the hierarchy equations are taken, that is, they are multiplied by products of $e_x$ and
$e_y$ and integrated over the angle $\theta$. This calculation leads to evolution equations for
density and momentum, however, split up for the different time scales. For example, there are
separate equations for $\partial_{t_0}\rho$ and for $\partial_{t_2}\rho$. Successively inserting
and partially solving the equations, and finally adding all pieces together, for example like
$\partial_t\rho= \partial_{t_0}\rho+\partial_{t_1}\rho+\partial_{t_2}\rho + \ldots$ ($\epsilon$ has
been set to one at this stage) leads to the desired hydrodynamic equations, Eqs. (94) and (130) in
Ref.~\citenum{ihle_16}.

The microscopic collision rules enter this procedure through the velocity moments of the collision
integral $C$, {\it i.e.} through quantities like $\langle e_x C_1\rangle$ or
$\langle e_x e_y C_2 \rangle$ with $\langle\ldots\rangle \equiv \int_0^{2\pi}\ldots {\rm d}\theta$.
For example, the former quantity is the $O(\epsilon)$ contribution of the following moment, 
\begin{eqnarray}
\nonumber
& &\langle e_x\,(C\circ f) \rangle=\langle v_0 \cos{\theta}\, (C\circ f) \rangle={2 v_0\over \eta} \sin{\eta\over 2} 
 \sum_{n=1}^{\infty}{{\rm e}^{-M}\over (n-1)!} \int {\rm d}\tilde{\theta_1}\ldots {\rm d}\tilde{\theta_n} \\
\nonumber         
& &\int_{\odot} {\rm d}{\bf x}_2\ldots {\rm d}{\bf x}_n
\cos{\Phi_1}\,
[f_0+\epsilon f_1({\bf x},\tilde{\theta}_1)+\epsilon^2 f_2({\bf x},\tilde{\theta}_1)]
[f_0+\epsilon f_1({\bf x}_2,\tilde{\theta}_2)+\epsilon^2 f_2({\bf x}_2,\tilde{\theta}_2)]\ldots \\
\label{EX_MOMENT1}
& &
\times[f_0+\epsilon f_1({\bf x}_n,\tilde{\theta}_n)+\epsilon^2 f_2({\bf x}_n,\tilde{\theta}_n)]+O(\epsilon^3)
\end{eqnarray}
and is defined as
\begin{equation}
	\langle e_x C_1\rangle = \lim_{\epsilon\to 0}
	{\partial\over \partial\epsilon}\langle e_x\, (C\circ f)\rangle 
\end{equation}
In these moments of $C\circ f$, a crucial approximation was made in Refs. \citenum{ihle_11}, 
\citenum{ihle_14_a} and \citenum{ihle_16}, that led to the formal absence of collisional contributions
to the transport coefficients. This approximation consists of neglecting spatial variations of the
distribution $f$ across the interaction circle. This issue comes up because the Enskog-like collision
term $C\circ f$ involves integrals with products of $f$ over the collision circle. Here, we abandon
this approximation which is not justified if the interaction radius is of the same order or larger
than the mean free path, {\it i.e.} $R \gtrsim \Lambda$.

Comparing Eqs.~(\ref{FOURIERB1}) and (\ref{FOURIERC1}) with (\ref{DEF_HD_FIELDS}) leads to the
identification of the Fourier coefficients $a_1$ and $b_1$  with the components of the momentum
density, ${\bf w}=(a_1, b_1)$. Now, inserting $f_0$ and $f_1$ from Eqs.~(\ref{FOURIERB1}) and
(\ref{FOURIERC1}) into Eq.~(\ref{EX_MOMENT1}), and performing the integrations yields 
\begin{eqnarray}
\nonumber
& &\langle e_x\,(C\circ f)=\epsilon\,
{4\over \eta} \sin{\eta\over 2}\sum_{n=1}^{\infty}
{{\rm e}^{-M}\over (n-1)!} K_C^1(n) \\
\label{URMOMENT}
& &\left[ M^{n-1} w_x({\bf x})+(n-1)M^{n-2}\rho({\bf x})
\int_{\odot}{\rm d}{\bf x}_2\, w_x({\bf x}_2)\right]
+O(\epsilon^2)	
\end{eqnarray}
with
\begin{equation}
	K_C^1(n) = {1\over (2\pi)^n} \int {\rm d}{\tilde{\theta}_1}\ldots {\rm d}{\tilde{\theta}_n}
	\cos{\Phi_1(\tilde{\theta_1},\ldots, \tilde{\theta_n})}\cos{\tilde{\theta_1}}
\end{equation}
The $n$-dimensional angular integral, $K_C^1$, has been evaluated before, see table I in Ref.
\citenum{ihle_16}.

Expanding the density and the $x$-component of the momentum density around ${\bf x}$ and decorating
every spatial gradient with a power of $\epsilon$ gives
\begin{eqnarray}
\label{EXPAND_RHO}
& &\rho({\bf x}_2)=\left[1+\epsilon (x_{2,\alpha}-x_{\alpha})\partial_{\alpha}+ 
\epsilon^2 (x_{2,\alpha}-x_{\alpha}) (x_{2,\beta}-x_{\beta}) \partial_{\alpha}\partial_{\beta}+\ldots\right]\,
\rho({\bf x})\\
\label{EXPAND_W}
& &w_x({\bf x}_2)=\left[1+\epsilon (x_{2,\alpha}-x_{\alpha})\partial_{\alpha}+ 
\epsilon^2 (x_{2,\alpha}-x_{\alpha}) (x_{2,\beta}-x_{\beta}) \partial_{\alpha}\partial_{\beta}+\ldots\right]\,
w_x({\bf x})\,.
\end{eqnarray}
Only the first term from Eq. (\ref{EXPAND_W}) will contribute to $\langle e_x C_1\rangle$ since the
gradient terms are higher order in $\epsilon$. Thus, we can replace $\int_{\odot}{\rm d}{\bf x}_2\, 
w_x({\bf x}_2)$ by $A w_x({\bf x})$ where $A=\pi R^2$ is the area of the collision circle. Inserting
the expansion~(\ref{EXPAND_RHO}) into the defining equation for $M$,
\begin{equation}
	M({\bf x})=\int_{\odot}\rho({\bf x}_2)\,d{\bf x}_2
\end{equation}
one finds $M({\bf x})=A\rho({\bf x})+O(\epsilon^2)$. Thus, $\rho({\bf x})$ can be approximated by
$M/A$ in Eq.~(\ref{URMOMENT}) if one only cares about the first order contribution $\langle e_x C_1
\rangle$. This result is identical to Eqs.~(38) and (39) in Ref.~\citenum{ihle_16},
\begin{equation}
	\langle e_x C_1\rangle = \lambda w_x({\bf x})
\end{equation}
with the factor $\lambda$ 
\begin{equation}
	\lambda \equiv {4\over \eta}\sin{\eta\over 2} {\rm e}^{-M}
	\sum_{n=1}^{\infty} {M^{n-1} n\over (n-1)!} K_C^1(n)\,.
	\label{DEF_LAM}
\end{equation}
This factor was discussed in detail in Ref.~\citenum{ihle_16} and it describes the ensemble-averaged
amplification of the momentum density. The threshold condition for the transition to collective
motion is given by $\lambda=1$ (assuming molecular chaos and a spatially homogeneous system). For
$M \gg 1$, Eq.~(\ref{DEF_LAM}) can be approximated as 
\begin{equation}
	\lambda \approx {1\over \eta}\sin\left({\eta\over 2}\right) \sqrt{(M+1)\pi}\,,
	\label{LAMBDA_LARGE_M}
\end{equation}
whereas for $M \ll 1$ one finds
\begin{equation}
	\lambda \approx {2\over \eta}\sin\left({\eta\over 2}\right) {1+4M/\pi+0.7872M^2+0.3M^3\over 1+M+M^2/2+M^3/6}\,.
	\label{LAMBDA_SMALL_M}
\end{equation}

Similar to the calculation above, we recalculated moments of the collision operator in second order
in $\epsilon$ such as $\langle e_x^2 C_2\rangle$ and $\langle e_x e_y C_2\rangle$ without the
approximation of large mean free path and again did not see any difference to previous results. Thus,
we conclude that, at least at a mean-field level, previous calculations of transport coefficients
that depend solely on moments of $C \circ f$ in linear and quadratic order in $\epsilon$ remain
correct at small mean free paths. However, in third order in $\epsilon$, additional terms arise that
were neglected previously in the large mean free path approximation. Consider the third order
contribution to the moment from Eq. (\ref{URMOMENT}),
\begin{equation}
\label{THIRD_ORDER_MOM}
	\langle e_x C_3\rangle=
	\lim_{\epsilon \to 0} {1\over 3!} {\partial^3\over\partial\epsilon^3}\langle e_x\, (C\circ f)\rangle
\end{equation}
which, according to Eq.~(\ref{URMOMENT}) contains an integration of the momentum density over the
collision circle, $\int_{\odot}{\rm d}{\bf x}_2\, w_x({\bf x}_2)$ where the expansion~(\ref{EXPAND_W})
is inserted, and the integration over the collision circle can be performed explicitly in every term of
the series. This calculation gives,
\begin{equation}
	\label{INT_EXPAND}
	\int_{\odot}{\rm d}{\bf x}_2\, w_x({\bf x}_2) = 
	A w_x({\bf x})+{\epsilon^2\over 2}\int_0^R r^3 {\rm d}r \int_0^{2\pi} {\rm d}\alpha \,\hat{n}_{\alpha}\hat{n}_{\beta}
	=A\left[1+{\epsilon^2 R^2 \over 8}\nabla^2+O(\epsilon^4)\right] w_x({\bf x}) 
\end{equation}
where $\hat{n}=(\hat{n}_x,\hat{n}_y)=(\cos{\alpha},\sin{\alpha})$ is the radial unit vector. Terms
with odd powers of $\epsilon$ disappear because of the symmetric (circular) shape of the collision
area.

Inserting Eq.~(\ref{INT_EXPAND}) into Eq.~(\ref{URMOMENT}) leads together with Eq.
(\ref{THIRD_ORDER_MOM}) to
\begin{equation}
	\langle e_x C_3\rangle=\Gamma w_x w^2+S(w_x a_2+ w_y b_2)+H\nabla^2 w_x
\end{equation}
where the coefficients $\Gamma$ and $S$ are given in Eqs. (61) and (62) of Ref.~\citenum{ihle_16},
and $a_2$ and $b_2$ are Fourier coefficients defined in Eq.~(\ref{FOURIERC1}). The new result of
the current paper is the third term whose coefficient $H$ is,
\begin{equation}
	H=
{R^2\, \sin{(\eta/2)}\over 2 \eta}
        \sum_{n=1}^{\infty} {{\rm e}^{-M}\over (n-1)!} M^n\, K_C^1(n+1)
        \label{H_DEF}
\end{equation}
We checked that relaxing the previous restriction on the mean free path only affects the moment
$\langle e_{\beta} C_3\rangle$ and does not impact other relevant moments, at least in a third-order
CE expansion. In order to obtain improved transport coefficients, it therefore suffices to formally
replace all occurrences of $\Gamma w_{\beta} w^2$ by $\Gamma w_{\beta} w^2+H\nabla^2 w_{\beta}$ in
the calculations of Ref. \citenum{ihle_16} after Eq. (111) of that paper. As a result of this
straightforward but technical exercise, we observed that, at least up to third order in $\epsilon$,
all transport coefficients except the viscosity remain unchanged. In particular, we found the novel
collisional contribution to the kinematic viscosity, $\nu_{\rm coll}$, as presented in
Eq.~(\ref{NU_COLL_ALL}) above.

The time scale $t_0$, introduced in Eq.~(\ref{GRAD_EXPAND}), is the fast convective time that is
associated with the Euler-equation (hence non-dissipative), and it measures the time, momentum is
convected due to a pressure gradient. The multitime scale expansion, Eq.~(\ref{GRAD_EXPAND}), allows 
that an approximation of a given order can vary rapidly with respect to one time scale but more
slowly with respect to another. Here, the density $\rho$ does not vary on the time scale $t_0$, 
{\it i.e.} $\partial_{t_0}\rho=0$. It turns out that the time scale $t_1$ is spurious and physically
irrelevant in the chosen vicinity to the transition threshold, where we assumed $1-\lambda = 
\mathcal{O}(\epsilon^2)$. This is because both hydrodynamic variables, density $\rho$ and momentum
density $\mathbf{w}$, do not change at all on this scale, so that $\partial_{t_1}\rho=0$ and
$\partial_{t_1}\mathbf{w}=0$. The timescale $t_1$ is only present in the current formalism due to
a systematic scaling Ansatz in powers of $\epsilon$, that is, for ``historical'' reasons. Finally,
the time scale $t_2$ is a slower relaxation time scale, which is associated with the viscous
processes that bring the system into its stationary state. For our chosen scaling, $1-\lambda =
\mathcal{O}(\epsilon^2)$, the local relaxation of momentum due to transferring it to and from
the environment (encoded in the alignment interaction and the angular noise), as well as the
nonlinear processes that lead to the cubic term $\propto w^2\mathbf{w}$ in the hydrodynamic
equations, also occur on the same time scale $t_2$ as the momentum diffusion. This can be seen
in Eq.~(114) of Ref.~\citenum{ihle_16}.

\section{Verification of the RP method for an MPCD fluid}
\label{sec:appendixB}
\label{sec:mpcd}
To validate our approach and the implementation of the shear algorithm, we used the RP method to
compute the shear viscosity of an MPCD solvent in two dimensions \cite{malevanets_99, malevanets_00,
gompper_08, kapral_08, howard_18, howard_19}. MPCD is a particle-based mesoscale technique, often
used for simulating the dynamics of complex fluids such as polymeric suspensions and blood flow. 
The general idea behind MPCD is to adopt a computationally inexpensive, coarse-grained solvent model
that faithfully reproduces the solvent-mediated hydrodynamic interactions. Similar as in the VM, the
motion of the MPCD particles is governed by alternating streaming and collision steps. During the
streaming step, the velocities of all fluid particles are updated according to Eq.~(\ref{STREAM}).
In the collision step, the fluid particles undergo stochastic collisions with particles in the same
quadratic collision cell, where the edge length of these cells, $a$, dictates the spatial resolution
of the hydrodynamic interactions \cite{huang:pre:2012}. Here, we employed both the stochastic
rotation dynamics (SRD) \cite{malevanets_99} and the Andersen thermostat (AT) collision rule
\cite{allahyarov:pre:2002}. For both variants of the MPCD algorithm, analytic expressions and
previous numeric calculations for the transport coefficients are readily available in the literature
\cite{tuzel_03, kikuchi_03, ihle_05, pooley_05}.

\begin{figure}[htbp]
	\includegraphics[width=\figwidth]{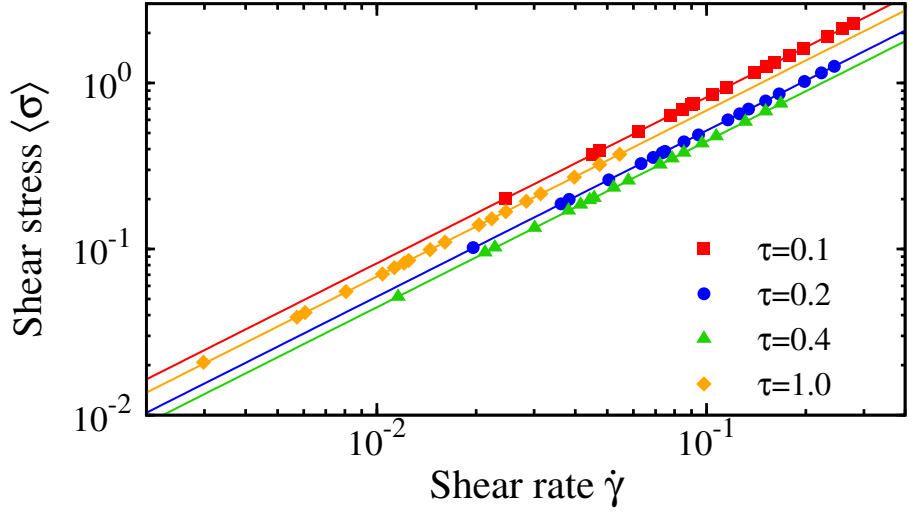}
	\caption{Shear stress, $\la \sigma \ra$, {\it vs.} shear rate, $\dot{\gamma}$, for the MPCD-AT
	algorithm with $\rho=10a^{-2}$ and $k_{\rm B} T=1$ at various time steps $\tau$. The symbols
	correspond to simulation data, while the lines are linear fits.}
	\label{fig:mpcdShearStress}
\end{figure}

We achieved isothermal conditions in the MPCD-SRD simulation by employing a Monte Carlo style
thermostat \cite{hecht_05, gompper_08}, which correctly conserves the local momentum in each
collision cell and reproduces the desired Maxwell velocity distribution. In the MPCD-AT simulations,
thermalization was achieved directly through the collision step. Galilean invariance was restored
by applying a random shift of the collision cells before every collision step \cite{ihle_01}.
In all MPCD simulations, the particle mass $m$ was set to unity, and a temperature of $k_{\rm B} T=1$
was used. Simulations were conducted in a quadratic simulation box with $L_x=L_y=16a$ and
periodic boundary conditions in all directions. A particle number density of $\rho=10a^{-2}$
has been used throughout. For the SRD rule, we set the collision angle to $\alpha = 110^\circ$. We
determined the shear viscosity $\eta$ of the MPCD fluids for time steps $\tau = 0.1$, $0.2$, $0.4$
and $1.0$, by conducting multiple simulations at different average shear stress $\la \sigma \ra$.
Figure~\ref{fig:mpcdShearStress} shows $\la \sigma \ra$ {\it vs.} the measured shear rate,
$\dot{\gamma}$ for the MPCD-AT simulations (the MPCD-SRD results are qualitatively similar),
demonstrating that the MPCD fluid behaves like a Newtonian liquid, as expected. From these data,
the shear viscosity can then be computed as $\nu = \la \sigma \ra /\dot{\gamma}$.

\begin{figure}[htbp]
	\includegraphics[width=\figwidth]{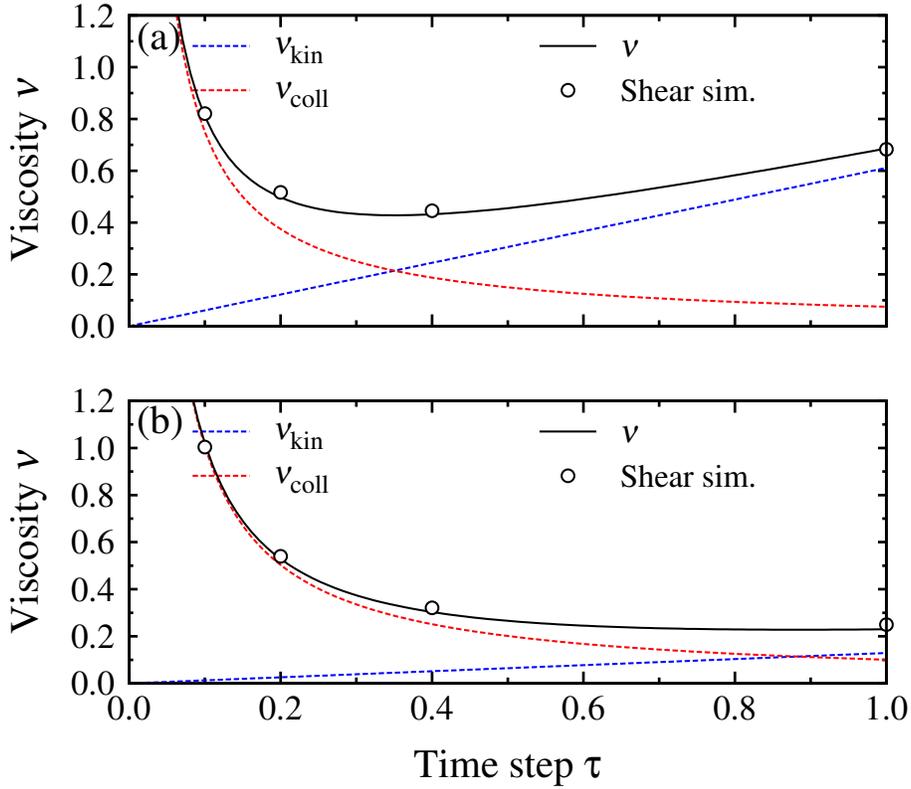}
	\caption{Viscosity of an MPCD fluid at $\rho = 10a^{-2}$ and $k_B T=1$ using the (a) AT collision
	scheme,	and the (b) SRD variant with $\alpha = 110^\circ$. The lines correspond to the theoretical
	prediction, while symbols show the simulation results.}
	\label{fig:mpcdViscosity}
\end{figure}

Figure~\ref{fig:mpcdViscosity} shows $\nu$ as a function of $\tau$ compared to the theoretical
prediction for the MPCD-AT and MPCD-SRD algorithm. In both cases, the shear viscosity of the fluid,
$\nu=\nu_{\rm coll}+\nu_{\rm kin}$, is dominated at small $\tau$ by the collisional contribution,
$\nu_{\rm coll}$. However, as $\tau$ is increased (and thus the mean free path of the particles,
$\tau \sqrt{k_B T/m}$, becomes larger), particle collisions become less important, and the shear
viscosity of the fluid is dominated by the kinetic contribution, $\nu_{\rm kin}$, instead. The
viscosity computed from the shear simulations follows these trends perfectly, and we achieved
quantitative agreement with the theoretical expressions  within $3\,\%$. This small difference of a
few percent between theory and simulation is in the same range of errors which were observed
 previously by using other methods such as GK relations \cite{pooley_05, gompper_08, ihle_05}.

\begin{thebibliography}{99}
\bibitem{foot1} A kinetic theory that goes beyond the mean field assumption of molecular chaos can be found in Ref. \citenum{chou_15}.
\end{thebibliography}
\end{document}